\documentclass[10pt,preprint]{aastex}  

\newcommand{\myemail}{takahasi@phyas.aichi-edu.ac.jp}

\slugcomment{accepted for publication in ApJ}

\shorttitle{Shocks in Trans-Magnetosonic Accretion Flows} % onto a Black Hole 
\shortauthors{Takahashi et al.}

\begin{document}

%\date{ \today } % (June 6, 2003) ...  

%% LaTeX will automatically break titles if they run longer than
%% one line. However, you may use \\ to force a line break if
%% you desire.

\title{ STANDING SHOCKS IN TRANS-MAGNETOSONIC 
        ACCRETION FLOWS ONTO A BLACK HOLE}

\author{Masaaki Takahashi, Junya Goto}
\affil{Department of Physics and Astronomy, Aichi University of Education, \\ 
       Kariya, Aichi 448-8542, Japan; \\
       \myemail, jgoto@hst.phyas.aichi-edu.ac.jp}
\and
\author{Keigo Fukumura,  Darrell Rilett,  and  Sachiko Tsuruta}
\affil{Department of Physics, Montana State University,
       Bozeman, MT 59717-3840;
       fukumura@physics.montana.edu, 
       rilett@physics.montana.edu, 
       tsuruta@physics.montana.edu}

\begin{abstract}
 Fast and slow magnetosonic shock formation is presented for
 stationary and axisymmetric magnetohydrodynamical (MHD) accretion 
 flows onto a black hole. The shocked black hole accretion solution must
 pass through magnetosonic points at some locations outside and inside
 the shock location. We analyze critical conditions at the magnetosonic
 points and the shock conditions. Then, we show the restrictions on the
 flow parameters for strong shocks. We also show that a very hot shocked
 plasma is obtained for a very high-energy inflow with small number
 density. Such a MHD shock can appear very close to the event horizon,
 and can be expected as a source of high-energy emissions. 
% Examples of these magnetosonic shocks are presented.    
 Examples of shocked MHD accretion flows are presented in the
 Schwarzschild case.  
\end{abstract}


\section{Introduction}

 In connection with the activity of active galactic nuclei (AGNs),  
 compact X-ray sources 
 and inner engines of gamma-ray bursts (GRBs), %[add]
 we consider accreting plasmas onto a black hole.  
 The observed huge energy outputs are mainly originated from the 
 gravitational energy released from infalling matters. When a black hole   
 is rapidly rotating in an external magnetic field, we can also expect 
 the release of the rotational energy of the black hole by the
 electromagnetic interaction between the black hole's spin and the
 magnetic field (see \cite{BZ77} in the force-free limit;
 \cite{TNTT90}, hereafter TNTT90; \cite{Hirotani92} for ideal MHD flow).   
 The global magnetic field around a black hole should be enhanced by
 infalling magnetized plasmas, and the magnetic field generated by the
 dynamo motion of the surrounding plasma can also interact with the spin
 of the black hole.   
 The ``black hole magnetosphere'' system is composed of a rotating black
 hole, surrounding plasmas and magnetic fields, and considered as a
 central engine of AGN 
 and GRB   %[add]
 activities.  It would output energy in the form
 of high-energy radiation (X-rays and $\gamma$-rays) and
 highly-accelerated plasma outflows (relativistic jets and winds). 
 In both energy output processes, the study of plasma behavior in the
 magnetosphere is the key to various problems. This situation is very
 similar to the activity of the solar magnetosphere (i.e., the emission
 of X-rays and the generation of solar wind).  
 In this paper, we investigate the basic plasma physics of the vicinity
 very close to a black hole including the magnetosphere.

 The magnetosphere is assumed stationary and axisymmetric, and the ideal
 MHD approximation is assumed. Then, the MHD plasmas stream along
 magnetic field lines in the magnetosphere, and five field aligned 
 parameters exist; that is, the total energy, total angular momentum, 
 number flux per magnetic flux and entropy of the MHD flow, and the
 angular velocity of the magnetic field line (see next section).   
 The flows accelerated by gravity pass through magnetosonic 
 points, where the poloidal component of the flow velocity becomes the
 magnetosonic wave speed. To fall into the black hole, the accreting
 plasma ejected from the plasma source with low-velocity must pass
 through the fast and slow magnetosonic points and the Alfv\'{e}n point,
 which are singular points in the basic equations. The physical flow
 solution along a magnetic field line is restricted by the regularity
 conditions at these singular points \citep{Takahashi02a}. That is, the
 conditions restrict the possible ranges of physical flow parameters.

 Hereafter, we will discuss the MHD shock in accretion onto a black hole.  
 Relativistic MHD shocks are discussed by, e.g., \cite{Ardavan76} and
 \cite{Lichnerowicz67,Lichnerowicz76}.  We apply their formalism to a
 black hole magnetosphere in Kerr spacetime (for the slow magnetosonic
 shock, see \cite{TRFT02}, hereafter TRFT02;
 \cite{Takahashi02b,Rilett03}).   %[add]
 The trans-magnetosonic accretion flow, which is ejected from a plasma 
 source and passes though the first magnetosonic point, results in the
 shock on the way to the event horizon, and the postshock
 sub-magnetosonic flow passes through the second magnetosonic point
 (located inside the shock front) again. 
 Such a MHD shock accretion solution must satisfy the critical conditions
 at both the first and second magnetosonic points and the shock
 conditions. Thus, we must adjust the five parameters of the flow to the  
 physically acceptable shocked 
% multi-magnetosonic accretion solution 
 accretion solution with multiple magnetosonic points 
 by considering both the critical conditions at the magnetosonic points
 and the jump conditions at the shock front.

 In a plausible black hole magnetosphere, global magnetic field lines
 are generated by the surrounding plasma (e.g., an equatorial disk and
 its corona).  We expect that (1) the magnetic field lines shaped like
 loops would be distributed around the inner-edge region of the disk,  
 (2) the disk surface and the black hole are connected by the magnetic
 field lines, where on the disk side of the field lines the footpoints
 can be anchored to the disk surface at the location of several times
 the inner-edge radii and on the black hole side the field lines can
 connect the event horizon from its pole to equator, and 
 (3) the open magnetic field lines connecting the outer region of the
 disk surface are extended to distant regions 
 \citep{Putten99,TT01,Li02,Uzdensky05}.   
%-[modified]->  
 The ingoing (or outgoing) flow is ejected with non-zero velocity from
 the footpoint on the disk surface, which is a plasma source and is
 called the ``injection point''. 
 Then, along the disk--black hole connecting magnetic field lines,  
 the MHD ingoing flows fall into the black hole, while the
 outgoing winds would stream along the open magnetic field lines 
 and the loop-like magnetic field lines would enclose static plasmas. 
 (For a high accretion rate accretion disk system, the plasma
 can fall into the black hole radially from the disk's inner-edge. In
 this case, the loop-like configuration of the magnetic field lines
 would disappear.)  By considering the disk--black hole magnetic field
 lines, non-equatorial accretion flow that falls into the black hole
 from the high-latitude region is possible. If a MHD shock arises on the
 non-equatorial inflows and forms a very hot plasma region by the shock,
 we can expect high-energy emissions in the magnetosphere, which would
 be distinct from the emissions from the equatorial accretion disk.   
 Such high energy emissions from the hot plasma would directly (or
 indirectly) carry informations for the strong gravitational field to us.  
%--

 In our numerical demonstrations for MHD shocked ingoing flows, we
 should solve the realistic magnetic stream function \citep{NTT91} for
 magnetic field lines connecting the disk and black hole. Note that
 at the shock front magnetic field lines bend toward toroidal and/or
 poloidal directions. However, in general, it is hard to solve
 self-consistently the magnetic structure of the magnetosphere.  
 So, for simplicity, without a realistic magnetic stream function, 
 we would only discuss the ingoing trans-magnetosonic flows on a 
 {\it conical} magnetic stream function. We assume that at the shock
 front the magnetic field lines bend only to the toroidal direction.  
 This assumption would be valid near the black hole (at least inside the
 inner-light surface). Near the injection point, of course, the conical
 magnetic fields would not be realistic. This is because the conical
 magnetic field lines do not connect to the equatorial disk surface;
 that is, we cannot define the footpoints. (Note that we can consider
 coronal gases distributed above the disk surface as a plasma source. 
 In this case, conical magnetic field may be probable.) 
% This is because the injection points cannot be determined in the
% conical field model, but under a certain realistic magnetic field model  
% configuration would locate near the equatorial disk region. 
 Furthermore, in the demonstrations in \S 4, we only treat the inflow
 streaming near the equatorial plane. However, we can expect that the
 qualitative picture is not drastically changed along the magnetic field
 lines (2) where we leave the equatorial region.

 In this paper, we discuss the general relativistic effects on the
 streaming MHD plasma in the black hole magnetosphere:  
 trans-magnetosonic accretion and MHD shock formation. 
 The accreting flow must be super-magnetosonic at the horizon, and the
 flow injected from the plasma source with low-velocity must pass
 through the magnetosonic points (the slow magnetosonic point, the
 Alfv\'en point, and fast magnetosonic point). 
%[cut] If the MHD shock is generated on the way to the event  
%  horizon, a very hot plasma region can be generated near the event
%  horizon, and high energy emissions from the hot plasma may directly (or
%  indirectly) carry informations for the strong gravitational field to us.  
 In \S 2, we present the general relativistic MHD flows. We introduce
 the field-aligned flow parameters and trans-magnetosonic solutions. 
 The MHD shock formation is studied and the shocked accretion solutions
 are presented in \S 3. We explore various types of shocked solutions.
 The properties of the MHD shock is discussed in \S 4.  
%-[modified]->  
 We can obtain a very hot plasma region near the event horizon 
 for the MHD shock of a high-energy inflow with small number density,  
 where the energy conversion from the kinetic energy to the magnetic
 energy is restricted because of the black hole boundary conditions on
 the toroidal magnetic field. 
 Summary and conclusions are given in \S 5.   
%--

\section{ Critical Conditions for Trans-Magnetosonic Flows }

 We consider stationary and axisymmetric ideal MHD accretion flows in
 Kerr geometry. In Boyer-Lindquist coordinates with the $c=G=1$
 unit, the metric of a Kerr black hole is given by   
\begin{eqnarray}
   ds^2 &=& \left( 1-\frac{2mr}{\Sigma} \right) dt^2 
        + \frac{4amr\sin^2\theta}{\Sigma} dt d\phi \\
       & &  - \frac{A\sin^2\theta}{\Sigma} d\phi^2
            - \frac{\Sigma}{\Delta} dr^2 - \Sigma d\theta^2  \ ,  \nonumber 
\end{eqnarray}
 where $\Delta \equiv r^2 -2mr +a^2 $, $\Sigma \equiv r^2 +a^2\cos^2\theta$,
 $A\equiv (r^2+a^2)^2-a^2\Delta \sin^2\theta$, 
 and  $m$ and $a$ denote the mass and angular momentum per unit mass 
 of the black hole, respectively. 
 The ideal MHD condition is $u^\beta F_{\alpha\beta}=0$, the particle
 conservation law is $(nu^\alpha)_{;\alpha} = 0$, where $u^\alpha$ is
 the fluid 4-velocity, $n$ is the number density of the plasma, and
 $F_{\mu\nu} = A_{\nu;\mu}-A_{\mu;\nu}$ is the electromagnetic tensor  
 ($A_\mu$ is a vector potential). The  equation of motion is
 $T^{\alpha\beta}_{;\beta}=0$. The energy-momentum tensor is given by   
\begin{equation}
 T^{\alpha\beta} = \ n\mu u^\alpha u^\beta -Pg^{\alpha\beta} 
 + \frac{1}{4\pi}\left( F^\alpha_{\ \lambda} F^{\lambda\beta}
 + \frac{1}{4} g^{\alpha\beta}F^2 \right) \ ,              \label{eq:emt}
\end{equation}
 where $\mu =(\rho+P)/n = m_{\rm part}+(P/n)\Gamma/(\Gamma-1)$ is the
 relativistic enthalpy, $P$ is the gas pressure, $\rho$ is the total
 energy density, $\Gamma$ is the adiabatic index, $m_{\rm part}$ is 
 the particle's mass, and $F^2\equiv F_{\mu\nu}F^{\mu\nu}$.  
 The electromagnetic field tensor $F_{\alpha\beta}$ satisfies Maxwell's
 equations, ${F^{\alpha\beta}}_{;\beta}=-4\pi j^\alpha$ and 
 ${{^\ast F}^{\alpha\beta}}_{;\beta}=0$, where 
 ${^\ast F}_{\alpha\beta} \equiv (1/2)\sqrt{-g} 
  \epsilon_{\alpha\beta\gamma\delta} F^{\gamma\delta}$ is the tensor
 dual to $F_{\alpha\beta}$ and $j^{\alpha}$ is the  electric current
 density. The determinant of the metric is $g$, and $\sqrt{-g} =  
 \Sigma\sin\theta$.  We also assume the relativistic polytropic
 relation $P=K\rho_0^\Gamma $ \citep{Tooper65}, 
 where $\rho_0 = nm_{\rm part}$ is the rest mass density, which is
 related to the total energy density by   
 $ \rho = \rho_0 + P/(\Gamma-1) $,   %% = \ \rho_0+{\hat n}P 
 and $K$ is a constant along the stream line for an ideal gas. 
 The magnetic and electric fields seen by a distant observer, which 
 are expressed in the Boyer-Lindquist coordinates, are defined by 
 $B_\alpha \equiv {^\ast F_{\alpha\beta}} k^\beta$ 
 and $E_\alpha = F_{\alpha\beta}k^\beta$, 
 where   $k^\alpha=(1,0,0,0)$ is the time-like Killing vector.

 In a stationary and axisymmetric magnetosphere, we can define
 magnetic field lines as $\Psi(r,\theta) = $ constant lines, where 
 $\Psi (=A_\phi)$ is the magnetic stream function. The plasma streams 
 along the magnetic field line with five constants of motions 
 \cite[see][]{Camenzind86a}:  the angular velocity of the field lines, 
 $\Omega_F(\Psi) = - F_{tr}/F_{\phi r} = - F_{t\theta}/F_{\phi\theta}$,
 the particle number flux per unit magnetic flux, 
  $ \eta(\Psi) = -n u^r G_t/B^r = -n u^\theta G_t/B^\theta 
   = n u^t(\Omega-\Omega_F)\rho_w^2 / B_\phi $, 
 where  $ G_t \equiv g_{tt}+g_{t\phi}\Omega_F $ and 
 $\Omega\equiv u^\phi/u^t$ is the angular velocity of the plasma,  
 the total energy of the magnetized flow, 
 $ E(\Psi) = \mu u_t -\Omega_F B_\phi /(4\pi \eta)$, 
 the total angular momentum, 
 $ L(\Psi) = -\mu u_\phi - B_\phi/(4\pi\eta)$, and 
 the entropy $S(\Psi)$, which is related to $K$. 
 Then, we also find the relations 
 $ E_r = \sqrt{-g}\, \Omega_F B^\theta / G_t $, 
 $ E_\theta = - \sqrt{-g}\, \Omega_F B^r / G_t $ and 
 $ B_t = E_t = E_\phi = 0 $. 
 From the poloidal components of the equation of motion with five field 
 aligned parameters, we can derive the general relativistic Bernoulli
 equation (the poloidal equation)
 \cite[][TNTT90]{Camenzind86b,Camenzind89} and the general relativistic 
 Grad-Shafranov equation (the trans-field equation)
 \citep{Camenzind87,NTT91}.

 The poloidal equation can be expressed by (see, e.g., TNTT90)
\begin{equation}
 \left( \frac{\mu}{\mu_c} \right)^2 (1+u_p^2) 
 = \left( \frac{E}{\mu_c} \right)^2
   \left[ (\alpha-2M^2)f^2 -k \right] \ , \label{eq:pol}
\end{equation}
 where 
 $ \alpha \equiv g_{tt}+2g_{t\phi}\Omega_F+g_{\phi\phi}\Omega_F^2 $, 
 $ f \equiv -(G_\phi + G_t \tilde{L})/[\rho_w(M^2-\alpha)] $,     
 $ k \equiv (g_{\phi\phi} + 2g_{t\phi}\tilde{L} +g_{tt}\tilde{L}^2)/
            \rho_w^2 $, 
 $ G_\phi \equiv g_{t\phi}+g_{\phi\phi}\Omega_F  
   =  g_{\phi\phi}(\Omega_F-\omega) $, 
 $\tilde{L} \equiv L/E$ and 
 $\omega\equiv -g_{t\phi}/g_{\phi\phi}$. 
 The relativistic Alfv\'{e}n Mach number $M$ is defined by 
\begin{equation}
 M^2\equiv \frac{4\pi\mu n u_p^2}{B_p^2} 
         = \frac{4\pi\mu\eta u_p}{B_p}  \ ,  \label{eq:mach2}
\end{equation}
 where the poloidal component $u_p$ of the velocity is defined by 
 $u_p^2 \equiv u^A u_A$ ($A=r,\theta$), and the poloidal component 
 $B_p$ of the magnetic field seen by a distant (lab-frame)
 observer is defined by $B_p^2 \equiv -B^A B_A G_t^{-2}$.
 The toroidal component of the magnetic field 
 $ B_\phi = (\Delta/\Sigma)F_{\theta r} $ can be reduced to 
 $ B_\phi = -4\pi\eta E \rho_w f $, which is expressed in terms of
 the field-aligned flow's parameters and the Alfv\'{e}n Mach number.   
 The denominator of the function $f$ becomes zero when $M^2=\alpha$, 
 where the poloidal velocity $|u_p|$ equals the relativistic
 Alfv\'{e}n wave speed. To obtain the physical MHD flow, which transits 
 from sub-Alfv\'{e}nic to super-Alfv\'{e}nic, we must require that the 
 numerator also must be zero. Then, from this regularity condition of 
 the function $f$, we obtain $\tilde{L} = g_{\phi\phi}^{\rm A}
 (\omega_{\rm A}-\Omega_F)/G_{t{\rm A}}$ (TNTT90). Note that, 
 for given $\tilde{L}$ and $\Omega_F$, we can find one or two radii 
 satisfying the above condition for $\tilde{L}$. We will denote this
 radius $r=r_{\rm A}$, the ``Alfv\'{e}n radius'', while the point
 satisfying both $M^2=\alpha$ and $r=r_{\rm A}$ is called the
 ``Alfv\'{e}n point''.  The label `A' indicates quantities at the
 Alfv\'{e}n radii. Note that, even if $M^2\neq\alpha$, the function $f$
 can be zero at the Alfv\'{e}n radius (not the Alfv\'{e}n point); the
 toroidal magnetic field $B_\phi$ becomes zero, while at the Alfv\'{e}n
 point $B_\phi$ has a non-zero value. We will denote the zero toroidal
 field location on a magnetic field line, the ``anchor point'' 
 \citep{Punsly01}. For a streaming MHD plasma, $B_\phi$ changes the sign
 across the anchor point.

 In the case of a rotating black hole magnetosphere, there are two light 
 surfaces given by $\alpha=0$: $r=r_{\rm L}^{\rm in}(\Psi)$ and
 $r=r_{\rm L}^{\rm out}(\Psi)$, where the rotational velocity of the
 magnetic field line becomes the light velocity as seen by a distant
 observer. The distribution of light surfaces depends on
 $\Omega_F(\Psi)$,  $a$ and $\Psi(r,\theta)$. The inner light surface 
 distributes near the event horizon and encloses it. The plasma  
 source must be located between the inner and outer light surfaces.  
 For a slowly rotating black hole case, $0\leq\omega_H <\Omega_F$, where  
 $\omega_H$ is the angular velocity of the black hole, and for a counter 
 rotating black hole, $\Omega_F<0$, {\it two}\/ locations of the
 Alfv\'{e}n radii are possible along a magnetic field line: 
 $r=r_{\rm A}^{\rm in}(\Psi)$ and $r=r_{\rm A}^{\rm out}(\Psi)$. 
 These radii are always located between the inner and outer light
 surfaces. Corresponding to two Alfv\'{e}n radii, we can find two groups
 of the trans-Alfv\'{e}nic MHD flow solutions. 
 In one of these the poloidal velocity equals the Alfv\'{e}n wave speed
 at the inner Alfv\'{e}n radii (called the inner Alfv\'{e}n point),
 while in the other at the outer Alfv\'{e}n radii (called the outer
 Alfv\'{e}n point). ( Note that these definitions of the {\it inner} and
 {\it outer} Alfv\'{e}n points are slightly different from those in TNTT90.)  
 In these cases, there is the minimum values of  ${\tilde L}\Omega_F$  
 ($ \equiv {\tilde L}^{\ast}_\pm \Omega_F$), for which the Alfv\'{e}n
 points exist in the magnetosphere. There, 
 $\tilde{L}={\tilde L}^{\ast}_{+} (>0)$ for $\omega_{H}<\Omega_F$ and
 $\tilde{L}={\tilde L}^{\ast}_{-} (<0)$ for $\Omega_F<0$ satisfy the
 relation $d{\tilde L}/dr_{\rm A}=0$.   
 On the other hand, for a rapidly rotating black hole case  
 $0<\Omega_F<\omega_H$, the number of the Alfv\'{e}n point is only 
 {\it one}\/ between the two light surfaces for an arbitrary $\tilde L$
 value. Note that the Alfv\'en points are located on the light surface
 when $\tilde{L} \Omega_F=1$; this corresponds to the magnetically
 dominated limit case.

 The critical conditions at the fast and slow magnetosonic points 
 are given by the differential form of the poloidal equation
 (\ref{eq:pol}), which is expressed as : 
\begin{equation}
  (\ln u_p)' = {\cal N}/{\cal D}\ , 
\end{equation}
 where
\begin{eqnarray}
   {\cal N} &=& \left(\frac{E}{\mu}\right)^2 \left\{
                \left[{\cal R}(M^2-\alpha )C_{\rm sw}^2 
                + M^4{\cal A}^2 \right] (\ln B_p)'  \right.  \nonumber \\
            & & \left.   
                + \frac{1}{2}(1+C_{\rm sw}^2) \left[M^4(M^2-\alpha )k' 
                - {\cal Q}\alpha ' \right] \right\}    \ ,   
                                                      \label{eq:numer} \\
   {\cal D} &=& (M^2-\alpha )^2\left[(C_{\rm sw}^2-u_p^2)(M^2-\alpha ) 
                + (1+u_p^2) M^4 {\cal A}^2{\cal R}^{-1} \right] \  
                                                      \label{eq:domom}
\end{eqnarray}
 with 
 ${\cal A}^2 \equiv e^2 +\alpha k = f^2(M^2-\alpha)^2$, 
 ${\cal R}   \equiv \alpha e^2 - 2 e^2 M^2 - k M^4 $, 
 ${\cal Q}   \equiv \alpha e^2 - 3 e^2 M^2 - 2k M^4 $, 
 and $ e \equiv 1 - \tilde{L}\Omega_F $. The prime $(')$ denotes 
 $ \partial_r+(B^\theta/B^r) \partial_\theta $,   
 which is a derivative along a stream line.  
 The relativistic sound velocity $a_{\rm sw}$ is given by  
\begin{equation}
   a_{\rm sw}^2 \equiv \left(\frac{\partial\ln \mu}
                                  {\partial\ln n} \right)_{\rm ad} 
          = (\Gamma-1)\frac{\mu-\mu_{c}}{\mu} \ , 
\end{equation}
 and the sound four-velocity is given by 
 $C_{\rm sw} = a_{\rm sw}/\sqrt{ 1-a_{\rm sw}^2 }$.

 When the poloidal velocity equals the fast or slow magnetosonic 
 wave speed, the numerator ${\cal D}$ becomes zero. The regularity
 of the physical solution, to be satisfied by the condition 
 ${\cal N}=0$, is also required at this critical location. The fast and
 slow magnetosonic points have X-type (saddle-type: physical) or O-type 
 (center-type: unphysical) topology on the integral curves. 
 The condition ${\cal D}={\cal N}=0$ is expressed in terms of four
 field-aligned conserved quantities.   
 When the values of the three conserved parameters are given at the
 plasma injection point, the value of the remaining parameter is
 specified by the critical conditions. \cite{Takahashi02a} discussed
 this problem in the Kerr geometry; for example, for given flow
 parameter sets of $\tilde{L}$, $\Omega_F$ and $\zeta_{\rm cr}$, the
 $\eta$-$r_{\rm cr}$ and $E$-$r_{\rm cr}$ relations were presented,
 where $\zeta_{\rm cr} \equiv (a_{\rm sw}^2)_{\rm cr}$ is related to the 
 entropy and the index ``cr'' means the fast or slow magnetosonic point
 (to indicate the fast or slow magnetosonic point, we can replace ``cr''
 by ``F'' or ``S'').  
 The fast magnetosonic point can be located between two Alfv\'{e}n
 radii, when $( 0 < ) {\tilde L}^{\ast}_\pm \Omega_F < 
 {\tilde L}\Omega_F <  {\tilde L}^{\rm max/min}_\pm \Omega_F ( < 1 )$. 
 Here, ${\tilde L}^{\rm max/min}_\pm$ are defined as $\tilde{L}$
 satisfying the relations both $d[\alpha(r_{\rm A})]/dr_{\rm A}=0$ 
 and $\alpha(r_{\rm A})=0$ (where the inner and outer Alfv\'en radii 
 coincide); ${\tilde L}^{\rm max}_{+}$ is used for $\omega<\Omega$ and
 ${\tilde L}^{\rm min}_{-}$ is used for $\Omega_F<0$. 
 In this case, the {\it hydro-like}\/ accretion solution discussed by
 \cite{Takahashi00,Takahashi02a} is possible. We will call
 such a fast magnetosonic point the {\it middle}\/ fast magnetosonic
 point. The hydro-like solution passes through the outer Alfv\'{e}n
 point and the middle fast magnetosonic point, and then falls  into the
 black hole.   
 However, for a rapidly rotating (or counter-rotating) plasma with
 $|\tilde L|>|{\tilde L}^{\rm max/min}_\pm|$, the hydro-like solution   
 is forbidden and the middle fast magnetosonic point disappears. This is
 due to the efficient centrifugal barrier on the fluid, so that no
 solution of ${\cal D}={\cal N}=0$ for the fast magnetosonic point
 appears between the inner and outer Alfv\'{e}n points.  
 On the other hand, the {\it magneto-like}\/ solution is possible for 
 $\tilde{L}_\pm^{\ast}\Omega_F < \tilde{L}\Omega_F < 1$ with the
 condition $\eta<\eta^{\rm max}$. The magneto-like solution passes
 through the inner Alfv\'{e}n point and the {\it inner}\/ fast 
 magnetosonic point located between the event horizon and the inner
 Alfv\'{e}n point, and then falls into the black hole. In this case,
 the fluid part of the angular momentum transports to the magnetic part
 of the angular momentum along the ingoing flow solution. The maximum
 value $\eta_{\rm max}$ for the inner fast magnetosonic point exists for
 a hot MHD flow, while $\eta_{\rm max}\to \infty$ in a cold limit.  
 We can expect that the inner fast magnetosonic point disappears for a
 strong MHD shock with larger $\eta$; such a MHD shocked inflow without
 the inner fast magnetosonic point is unphysical because the radial
 velocity of MHD inflow solutions becomes zero at the event horizon,
 where the number density diverges.  
 The slow magnetosonic points are also obtained from the same regularity
 condition ${\cal D}={\cal N}=0$.  We can find the inner, middle and
 outer slow magnetosonic points for a suitable $\eta$-value, which are
 located in the ranges of 
 $r_{\rm L}<r_{\rm S}^{\rm in}<r_{\rm A}^{\rm in}$,  
 $r_{\rm A}^{\rm in}<r_{\rm S}^{\rm mid}<r_{\rm A}^{\rm out}$, 
 $r_{\rm A}^{\rm out}<r_{\rm S}^{\rm out}<r_{\rm L}^{\rm out}$,
 respectively.  For the appearance of the inner or outer slow
 magnetosonic point, there is the restriction on the $\eta$-value 
 \cite[see][]{Takahashi02a}.

 In summary, in the case of $\omega_H<\Omega_F$ or $\Omega_F<0$,  
 $\tilde L$ must have the value within the range of 
 ${\tilde L}^{\ast}_\pm \Omega_F < \tilde L \Omega_F < 1$ to obtain
 two possible locations of the Alfv\'{e}n points in the $u_p$-$r$ (or 
 $M^2$-$r$) plane.  The magneto-like accretion solution is possible 
 when $\eta<\eta^{\rm max}$ and is effective for the magnetically
 dominated flows. The hydro-like accretion solution is realized when
 ${\tilde L}^{\ast}_\pm \Omega_F < {\tilde L}\Omega_F <
  {\tilde L}^{\rm max/min}_\pm \Omega_F$, and it becomes effective for
 the hydrodynamically dominated flows. For these two types of solutions, 
 which also depend on $E$, $\eta$ and $S$, the allowable parameter
 ranges are distinct, so that a discontinuous transition can be expected
 if the flow parameters change in a secular time-scale.   
 Note that, corresponding to two locations of the Alfv\'en point, two
 clearly distinct types of accretion solutions exist; the magneto-like 
 and hydro-like accretion solutions. On the other hand, for a rapidly
 rotating black hole case, $0<\Omega_F<\omega_H$, only one Alfv\'{e}n
 point exists, so that the distinction between the magneto-like and 
 hydro-like solution is not clear.   
 In the next section, we will only treat the case of $\Omega_F>\omega_H$
 and mainly consider the transition from the hydro-like solution to the
 magneto-like solution through the standing shock formation.

\section{ MHD Shock formation in curved spacetime} %-- in Kerr geometry  

 When we consider the shock formation for accreting MHD flows onto the
 black hole, we must require a fast (or slow) multiple magnetosonic
 solution, and must apply the general relativistic MHD shock condition
 between two fast (or slow) magnetosonic points. As mentioned in the
 previous section, the critical conditions restrict the allowable ranges
 of physical flow parameters. In such parameter ranges, we should
 restrict still more the flow parameter ranges for multiple magnetosonic 
 {\it shocked}\/ accretion solutions.

 \subsection{ The jump conditions} 

 Here, we discuss the jump condition for MHD flows in Kerr geometry.  
 The particle number conservation, the energy momentum conservation and   
 the magnetic flux conservation across the shock in a relativistic MHD
 flow are \citep{Lichnerowicz67,Lichnerowicz76,AC88}
\begin{eqnarray}
 & &  [nu^\alpha]^{1}_{2} ~ \ell_\alpha = 0 \ ,             \label{eq:nc} \\
 & &  [T^{\alpha\beta}]^{1}_{2} ~ \ell_\alpha = 0 \ ,       \label{eq:ec} \\
 & &  [^\ast F^{\alpha\beta}]^{1}_{2} ~ \ell_\alpha =0  \ , \label{eq:B}  
  %%  [u^\alpha b^\beta - u^\beta b^\alpha]\ell_\alpha=0
  %%   (Maxwell equation) 
\end{eqnarray}  
 where $\ell^\alpha$ is the unit vector normal to the shock front, 
 which has only poloidal components for stationary and axisymmetric
 flows ($\ell^A \ell_A = -1$). 
 The symbol $[Z]^{1}_{2} \equiv Z_{1}- Z_{2}$ means the change of a
 certain quantity $Z$ across the shock located at the radius 
 $r=r_{\rm sh}(\Psi)$, where the indeces ``1'' and ``2'' mean the
 properties of the preshock and postshock flows just on the shock
 front, respectively.    
 From the conditions (\ref{eq:nc}) and (\ref{eq:B}), the number flux 
 across the shock front $U \equiv n u^\alpha \ell_\alpha$ and 
 the normal component of the magnetic field (seen by a distant observer)
 to the shock front $B_\perp \equiv B^\alpha\ell_\alpha$  remain
 unchanged; that is, $[U]^{1}_{2} = 0$ and $[B_\perp]^{1}_{2} = 0$. We
 can also obtain $[E_\parallel]^{1}_{2} = 0$, $[\Omega_F]^{1}_{2} = 0$
 and $[\eta]^{1}_{2} = 0$, where $E_\parallel \equiv \epsilon_{AB}E^A\ell^B$
 is the tangential component of the electric field $E^\alpha$. The
 quantity $\eta$ can be expressed as $\eta=-UG_t/B_\perp $.    
 From the condition (\ref{eq:ec}), the following vector remains
 unchanged across the shock: 
\begin{equation}
  W^\alpha \equiv \mu u^\alpha U - \tilde{P}\ell^\alpha 
   -\frac{\alpha}{4\pi}\left(\frac{B_\perp}{G_t}\right)
    \left(\frac{B^\alpha}{G_t}\right) \ ,                 \label{eq:W}
\end{equation}
 where $\tilde{P} \equiv P+(B^2/8\pi)$ and 
 $B^2 \equiv (F^2/2) = \alpha B_p^2 +B_\phi^2/\rho_w^2$ . 
 From the product $W^\alpha \ell_\alpha$, we obtain the relation
\begin{equation}
   [ (\mu/n)U^2 + \tilde{P}]^{1}_{2} = 0 \ ,            \label{eq:Well}
\end{equation}
 which can be reduced to 
\begin{equation}
   U^2 = \frac{ \tilde{P}_2 - \tilde{P}_1 }{ (\mu/n)_1-(\mu/n)_2} 
       = 4\pi\eta^2 \frac{\tilde{P}_2-\tilde{P}_1}{M^2_1-M^2_2} 
       > 0 \ .  
\end{equation}
 From the conditions $[W^t]^{1}_{2} = 0$ and $[W^\phi]^{1}_{2} = 0$, 
 we also obtain that the constants of motion $E$ and $L$ are continuous
 across the shock, while entropy $S$, which is the fifth constant of
 motion, is discontinuous across the shock (see TRFT02). Of course, for
 the physically realistic shock solution the entropy must
 increase. Then, we will introduce the entropy related mass flow rate
 per magnetic flux tube defined by \cite{Chakrabarti90}    
%%\footnote{  
%% ~~ NOTE: ~~
%% The (specific) entropy is defined by $S\equiv \ln(P/\rho^\Gamma)$ 
%% (see {\it Kato ~in the textbook of ``Plasma Physics''}). 
%%}
\begin{equation}
 \dot{\cal M} \equiv K^{N} \mu_{c}\eta  \ ,
\end{equation}
 where $N=1/(\Gamma-1)$ is the polytropic index. The quantity
 $\dot{\cal M}$ conserves for a shock-free flow, but it must increase 
 at the shock due to the entropy generation. 
 By using the definitions of the Alfv\'{e}n Mach number and the
 relativistic enthalpy, we can express the entropy-related accretion
 rate as a function of $r$ and $M^2$ with the conserved quantities
\begin{equation}
 \dot{\cal M} = \frac{M^2}{4\pi\mu\eta} 
 \left[ \frac{1}{1+N}\left( \frac{\mu}{\mu_{c}}-1 \right) \right]^{N} \ ,  
\end{equation}
 where from the poloidal equation the relativistic enthalpy $\mu$ can
 be expressed as 
\begin{equation}
   \left(\frac{\mu}{\mu_{c}}\right)^2 = \left(\frac{E}{\mu_{c}}\right)^2 
             \left[ (\alpha-2M^2) f^2 - k \right]
           - \left( \frac{B_pM^2}{4\pi\mu_{c}\eta} \right)^2  \ .
\end{equation}
 Thus, we can plot trans-fast MHD solutions as $\dot{\cal M}(r, M^2)=$
 constant ($>0$) curves on the $r$-$M^2$ plane.  
 The physically acceptable MHD shock must satisfy the condition 
 $0 \leq \dot{\cal M}_1 < \dot{\cal M}_2$ ; $\dot{\cal M}=0$ for a cold
 flow ($\mu=\mu_{c}$).  The flow in the $\dot{\cal M}(r,M^2)<0$ region
 is forbidden as a physical solution.

 The location of the standing shock front and shock properties depend on
 the field-aligned flow parameters.  To discuss the shock properties, we
 will introduce the following dimensionless parameters
 \citep[][TRFT02]{AC88}: 
\begin{eqnarray}
 q     &\equiv& \frac{B^\phi_2}{B^\phi_1} 
       \ = \ \frac{M^2_1-\alpha}{M^2_2-\alpha} \ , \\
 \xi   &\equiv& \frac{n_2u^t_2}{n_1u^t_1} 
        \ = \ \frac{M^2_1(e-hM^2_2)}{M^2_2(e-hM^2_1)}q \ ,
\end{eqnarray}
 where $ h \equiv g^{tt}( 1 - \tilde{L}\omega )$.   
 The parameter $q$ is the ratio of the toroidal magnetic field before to
 that after the shock, and $\xi$ is the compression ratio; these
 parameters are quantities seen by a distant observer, and include the 
 gravitational red-shift factor and Lorentz factor for the plasma
 motion.  We also define the plasma frame compression ratio $\lambda$ by   
\begin{equation}
 \lambda \equiv \frac{n_2}{n_1} 
    \ = \ \frac{\mu_2 M^2_1}{\mu_1 M^2_2} \ . 
\end{equation}
 Furthermore, we are interested in the jump of the magnetization rate and
 temperature of the plasma. 
 The magnetization parameter $\sigma$, which is defined as the ratio of
 the Poynting flux to the total mass-energy flux seen by a zero angular
 momentum observer (ZAMO), can be expressed as (TRFT02)
\begin{equation}
 \sigma = \frac{B_\phi G_\phi}{4\pi\eta\mu u^t \rho_w^2} 
       \ = \ -\frac{e-\alpha h}{e-M^2 h} \ . 
\end{equation}
 The temperature parameter is defined by 
\begin{equation}
 \Theta \equiv k_{\rm B} T/m_{\rm part} = K \rho_0^{1/N} 
             = (\mu/\mu_{c}-1)/(1+N)  \ , 
\end{equation}
 where $T$ is the temperature of the fluid and $k_{\rm B}$ is the Boltzmann
 constant.

 \subsection{ Shocked accretion solutions}

 Now, let us consider MHD accretion solutions with a fast/slow
 magnetosonic shock on the $r$-$M^2$ plane. First, we need to find 
 two solutions of the trans-magnetosonic flow (i.e., {\it trans-fast}\/
 or {\it trans-slow}\/ magnetosonic flows corresponding to the fast or
 slow magnetosonic shock) having the same values of $E$, $L$, $\eta$ 
 and $\Omega_F$. By the shock formation, we can connect these two
 trans-magnetosonic solutions as a multiple magnetosonic accretion
 solution. Then, the shocked accretion flow passes through the fast or
 slow magnetosonic points twice. Along each branch of the preshock and
 postshock solutions, the value of $\dot{\cal M}$ (or the entropy $S$)
 is different, $\dot{\cal M}_{\rm pre} \neq\dot{\cal M}_{\rm post}$,
 where the indexes ``pre'' and ``post'' indicate the preshock and
 postshock trans-magnetosonic accretion solution. The values of 
 $\dot{\cal M}_{\rm pre}$ and $\dot{\cal M}_{\rm post}$ are specified  
 by the critical condition at each fast or slow magnetosonic point for
 the preshock and postshock solutions; that is, 
 $\dot{\cal M}_{\rm pre}=\dot{\cal M}(r_{\rm cr1}, M^2_{\rm cr1})$
 and $\dot{\cal M}_{\rm post}=\dot{\cal M}(r_{\rm cr2}, M^2_{\rm cr2})$, 
 where the labels `cr1' and `cr2' indicate the first and second fast or
 slow magnetosonic points, respectively.

 When we require a MHD flow solution passing through the slow
 magnetosonic point {\sf S}, the Alfv\'{e}n point {\sf A} and the fast
 magnetosonic point {\sf F} (hereafter ``SAF-solution''), from the
 critical conditions at the fast and slow magnetosonic point, a
 relation between $E$, $L$, $\eta$ and $\Omega_F$ is also specified.
 So, a change of the value of one of the parameters requires the change
 of the values of the remaining parameters.  
 In the next section, to demonstrate fast magnetosonic shocked
 accretion solutions, we will treat the hydro-like SAF-solution as an
 upstream flow solution and the magneto-like solution as a downstream
 flow solution for the fast magnetosonic shock formation. When we
 demonstrate a slow magnetosonic shock formation, we will consider that
 the downstream flow solution is the magneto-like SAF-solution, while
 the upstream flow solution is trans-slow magnetosonic solution.  
 Then, we can restrict the values of $\dot{\cal M}_{\rm pre}$ and
 $\dot{\cal M}_{\rm post}$ for the MHD flow solution passing through 
 the first and second X-type fast magnetosonic (or slow magnetosonic)
 points, respectively.  
 To find such a SAF-magnetosonic accretion solution, for a given 
 parameter set of $\Omega_F$, $E$ and $L$ we can search for the
 acceptable value of $\eta$ by tuning the regularity conditions 
 (${\cal D}={\cal N}=0$) at both fast and slow magnetosonic points.  
 If a suitable parameter set is obtained, we will find at least two 
 X-type fast/slow magnetosonic points on the $r$-$M^2$ plane.

 Next, by applying the relativistic jump condition, we can find one (or
 two) shock location to give $\dot{\cal M}_2=\dot{\cal M}_{\rm post}$ 
 with the condition $\dot{\cal M}_1=\dot{\cal M}_{\rm pre}$.  
 We will demonstrate an accretion solution with the MHD shock. 
 The accretion solution with a shock is obtained on the $r$-$M^2$ plane  
 by plotting $\dot{\cal M}(r,M^2)=$ constant curves of the two
 trans-fast magnetosonic (or trans-slow magnetosonic) flow solutions and
 a vertical line connecting these solution curves at the shock location.

%\placefigure{fig:saf} %---------  Figure~1 --------------------

\begin{figure}[t]%[h]% ------------------------------------- Figure~1
   \epsscale{0.4}
   \plotone{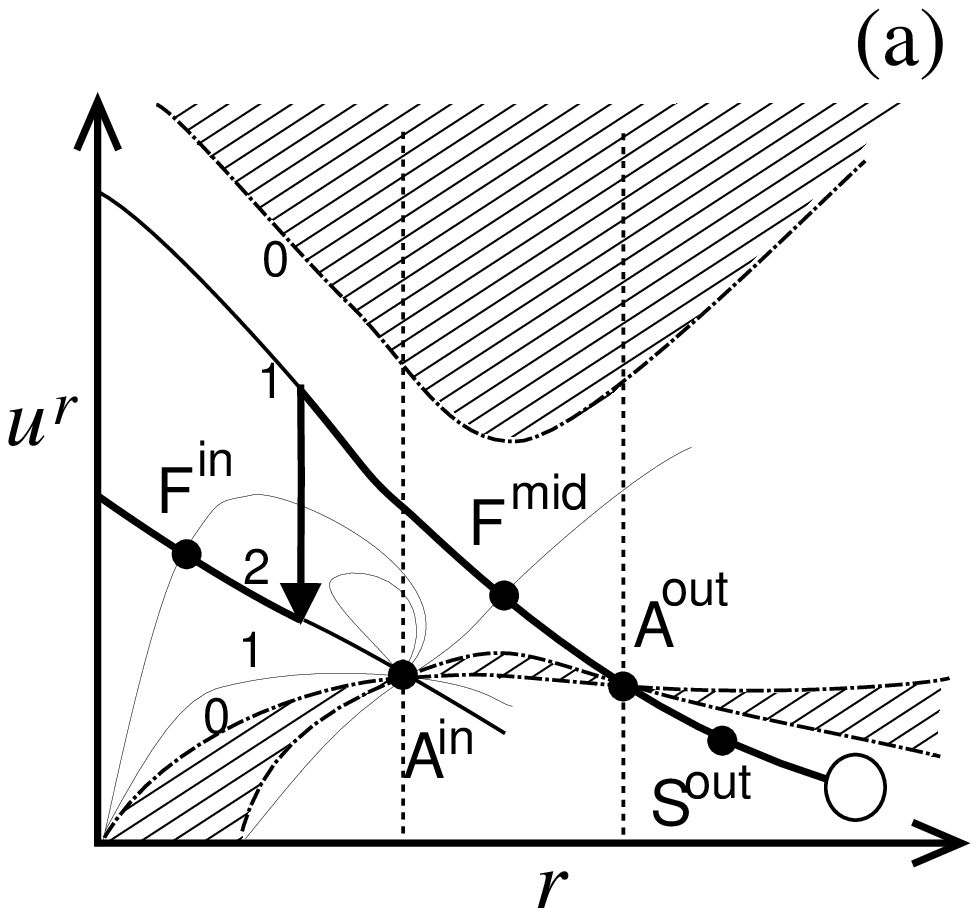}
   \plotone{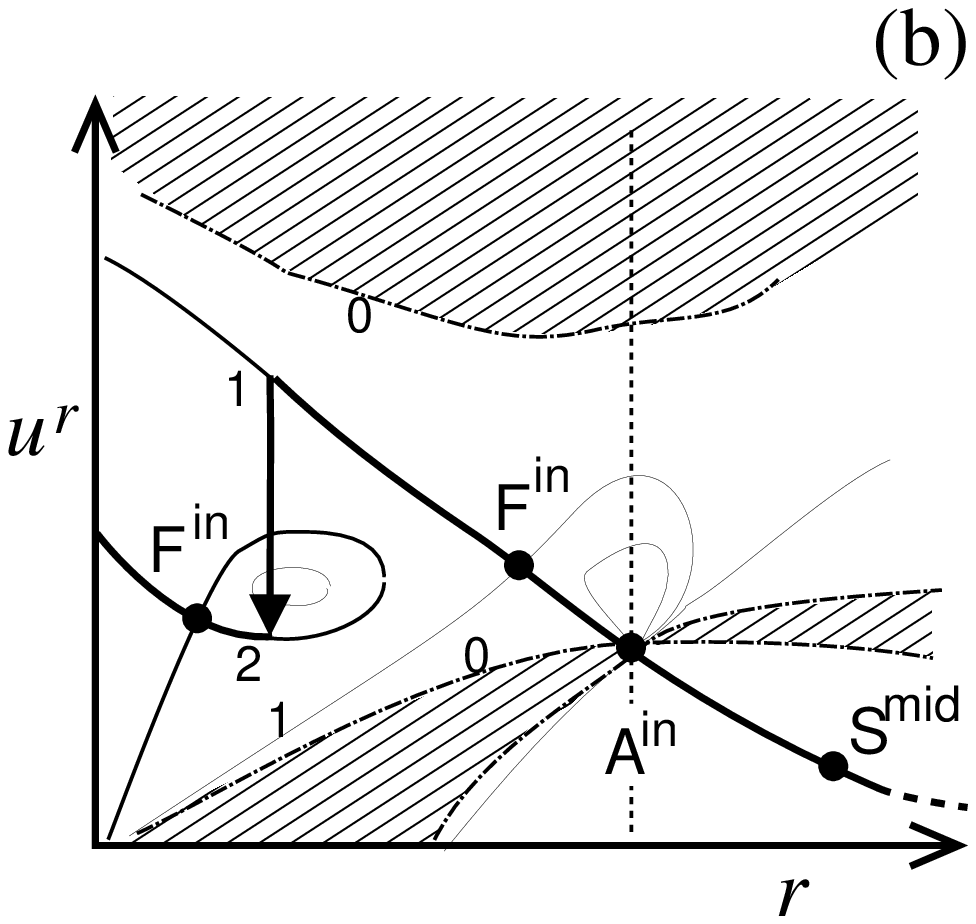}
  \caption{
   Schematic pictures of physically acceptable accretion solutions with
   the fast magnetosonic shock (thick curves and arrows). The upstream 
   SAF-solutions are (a) hydro-like and (b) magneto-like, and jump to
   the downstream trans-fast magnetosonic solution by the shock
   formation. The large circle in the Left panel (a) represents the
   plasma source, where the plasma is injected with a finite velocity. 
   In the Right panel (b), the downstream solution does not pass though
   the inner Alfv\'{e}n point, but connect to the event horizon with a
   finite Mach number. 
   The vertical dotted lines are the Alfv\'{e}n radii. The boundaries of 
   the hatched regions are $\dot{\cal{M}}=0$ curves and are labeled by
   ``0''. The $\dot{\cal{M}}$ value of the downstream curve with the
   label ``2'' is grater than that of the upstream with the label ``1''; 
   that is, the entropy of the flow is increased by the shock. 
   The other partially plotted thin curves that do not connect the
   plasma source are unphysical accretion solutions.       
  }
  \label{fig:saf}
\end{figure} %------

 Although the numbers (and their locations) of the fast and slow
 magnetosonic points depend on the flow parameters and the spin of the
 black hole \citep{Takahashi02a}, we can find the following cases of 
 the shocked MHD accretion solutions: two classes of fast magnetosonic
 shock accretion solutions,    
\begin{description}
 \item[~] FS-i~:~~ {\sf inj} 
	    $\to$ {\sf S}$^{\rm out}$ $\to$ {\sf A}$^{\rm out}$ 
	    $\to$ {\sf F}$^{\rm mid}$ $\to$ $<shock>$ 
	    $\to$ {\sf F}$^{\rm in}$ $\to$ {\sf H} \ . 
 \item[~] FS-ii~:~ {\sf inj} 
	    $\to$ {\sf S}$^{\rm mid}$ $\to$ {\sf A}$^{\rm in}$ 
	    $\to$ {\sf F}$^{\rm in}_{\rm outer}$ $\to$ $<shock>$ 
	    $\to$ {\sf F}$^{\rm in}_{\rm inner}$ $\to$ {\sf H} \ , 
\end{description}
and one class of slow magnetosonic shock accretion solution, 
\begin{description}
 \item[~] SS~:~~ {\sf inj} $\to$ {\sf S}$^{\rm out}$ $\to$ $<shock>$ 
	      $\to$ {\sf S}$^{\rm mid}$ $\to$ {\sf A}$^{\rm in}$ $\to$ 
              {\sf F}$^{\rm in}$ $\to$ {\sf H} \ . 
\end{description}
 In Figure~\ref{fig:saf}a (case FS-i), the upstream flow is a hydro-like 
 solution and the downstream postshock flow is a magneto-like solution.
 The shock formation is possible between these two trans-magnetosonic
 solutions, where the requirement $\dot{\cal M}_1 < \dot{\cal M}_2$ is
 satisfied. For the physically acceptable accretion solution, the
 postshock trans-magnetosonic solution must connect to the event horizon
 with nonzero four-velocity, $u^r_{\rm H}=$ finite. The upstream
 solution passes through the outer Alfv\'{e}n point ${\sf A}^{\rm out}$ 
 and the first X-type middle fast magnetosonic point ${\sf F}^{\rm mid}$.  
 After the fast magnetosonic shock, the downstream solution passes
 through the second X-type inner fast magnetosonic point ${\sf F}^{\rm
 in}$, and then falls into the black hole {\sf H}.    
 Figure~\ref{fig:saf}b (case FS-ii) shows that two X-type inner fast 
 magnetosonic points (inner ${\sf F}^{\rm in}$ and outer 
 ${\sf F}^{\rm in}$) locate inside the inner Alfv\'{e}n point. 
 The upstream magneto-like solution passes through the first X-type
 inner fast magnetosonic point ${\sf F}^{\rm in}_{\rm outer}$, after
 passing through the inner Alfv\'{e}n point ${\sf A}^{\rm in}$. The
 upstream solution can transit to the downstream trans-fast magnetosonic 
 accretion solution, which passes through the second X-type inner fast 
 magnetosonic point ${\sf F}^{\rm in}_{\rm inner}$ and connect to the 
 event horizon {\sf H}.   % with a finite (non-zero) value of $u^r$.    
 On the other hand, in the case of (SS), the upstream solution of the
 slow magnetosonic shock connects to the outer Alfv\'{e}n point, but
 before the Alfv\'{e}n point the upstream solution can jump to the
 downstream SAF-solution. This downstream solution passes though the 
 middle slow magnetosonic point ${\sf S}^{\rm mid}$, 
%[cut]  which is located between two Alfv\'{e}n radii, 
 the inner Alfv\'{e}n point 
 ${\sf A}^{\rm mid}$ and the inner fast magnetosonic point  
 ${\sf F}^{\rm in}$, and then falls into the black hole.

%\placefigure{fig:saf-x} %---------  Figure~2 --------------------

\begin{figure}[t]%[h]% ------------------------------------- Figure~2
   \epsscale{0.4}
   \plotone{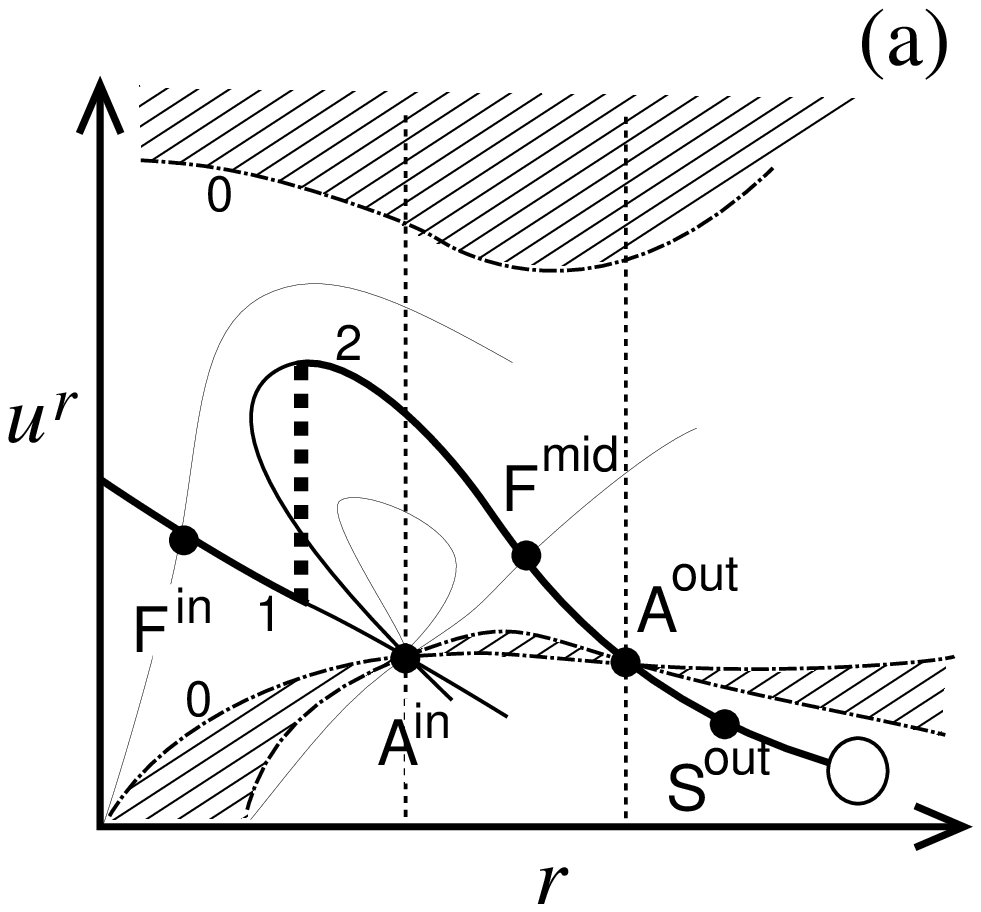}
   \plotone{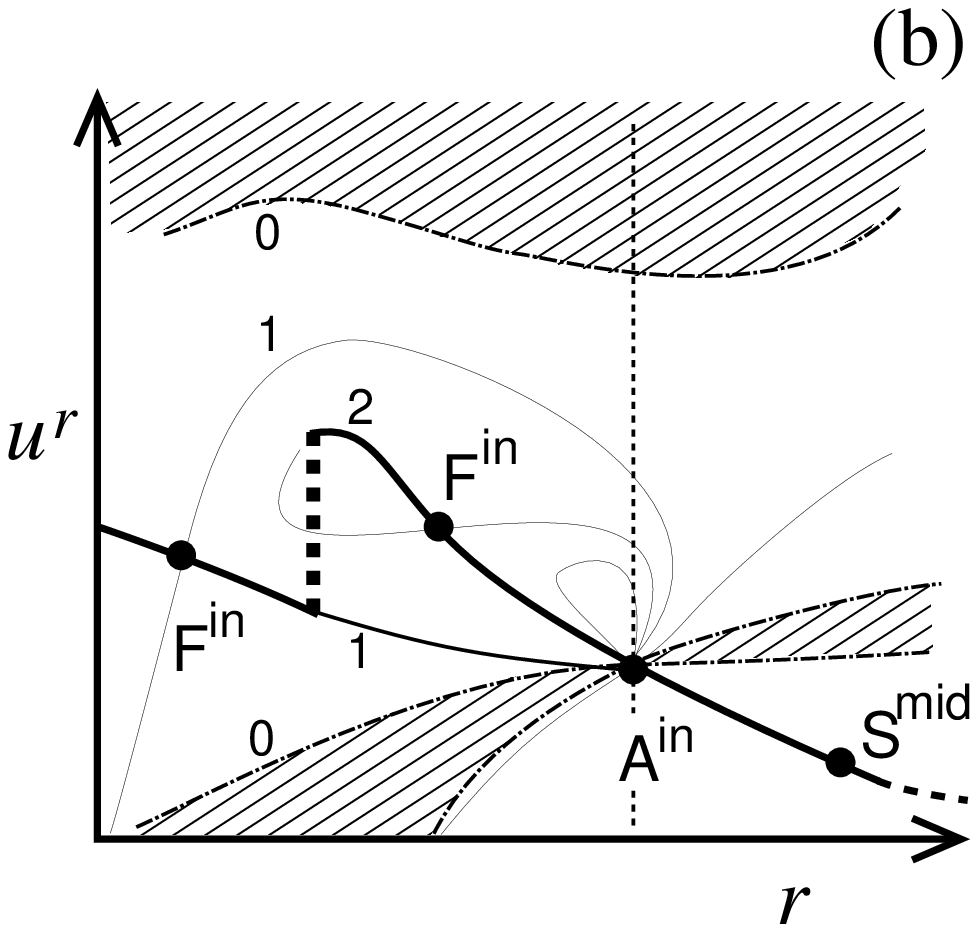}
  \caption{
   Schematic pictures of unrealistic accretion solutions. The upstream
   solutions are SAF-solutions, but are broken before the event
   horizon. In this case, the entropy (or $\dot{\cal{M}}$) of the
   upstream flow labeled by ``2'' is grater than that of the downstream
   labeled by ``1'', so that the shock formation is forbidden  (thick
   broken lines mean the disconnection of the upstream and downstream
   solutions). If the downstream curve is SAF-solution and connects to
   the plasma source and the event horizon, it is a physical accretion
   solution without a shock.   
  }
  \label{fig:saf-x}
\end{figure} %------

 One may consider that an upstream accretion solution does not need to
 connect to the event horizon {\sf H}. Such a solution is of course 
 unphysical as an accretion solution, but by the shock formation this 
 upstream solution may transit to a downstream physical trans-fast MHD
 accretion solution that connects to the horizon with a finite
 (non-zero) four-velocity. Although the upstream SAF-solution has  
 $\dot{\cal M}=\dot{\cal M}_{\rm pre}$ and the downstream trans-fast  
 MHD solution has $\dot{\cal M}=\dot{\cal M}_{\rm post}$, the
 $\dot{\cal M}=\dot{\cal M}_{\rm pre}$ curve makes a loop or a  
 double-loop on the $r$-$M^2$ plane by connecting to the inner
 Alfv\'{e}n point (and then it is disconnected to the event horizon) 
 as shown in Figs.~\ref{fig:saf-x}a and~\ref{fig:saf-x}b, while the
 $\dot{\cal M}=\dot{\cal M}_{\rm post}$ 
 curves enclose this loop and one of them connects to {\sf H}.    
 However, by considering the entropy (or $\dot{\cal M}$) distribution,  
 we conclude that the entropy of the disconnected upstream solution is
 greater than that of the downstream trans-fast MHD solution; that is,
 $\dot{\cal M}_{\rm pre}>\dot{\cal M}_{\rm post}$. So the transition
 from the upstream to downstream solution is forbidden.   
 Thus, for the fast magnetosonic shock formation, both upstream and
 downstream trans-fast accretion solutions need to connect to the event
 horizon with non-zero velocity. 
 We should note that even if two trans-magnetosonic accretion solutions
 exist for a set of values of the field-aligned quantities, no MHD shock
 may generate under such values. That is, we may find a MHD shock of  
 $\dot{\cal M}_1>\dot{\cal M}_2$, but such a solution is unphysical.

\section{ Numerical Results and Discussions }

 Now we show class (FS-i) fast magnetosonic shock solutions, where a
 cold ($\Theta\ll 1$) preshock flow is assumed. The solutions are
 given for the equatorial plane ($\theta=\pi/2$) in the Schwarzschild 
 spacetime ($a=0$).  
 For simplicity, we set the shock normal to the downstream flow; that
 is, $\ell^\alpha=(0,\ell^r,0,0)$.  Then, we obtain the relations 
 $B^r_1=B^r_2=B_\perp$ and $E^\theta_1=E^\theta_2=E_\parallel$.
 Furthermore, 
 without the trans-field equation  %[add]
 the poloidal magnetic field is assumed to be the radial
 configuration denoted by $ B^r = B_{0} G_t/\Sigma $ and $ B^\theta=0 $
 ( that is, $\hat{B_p}\equiv B_p/B_0 = 1/\sqrt{\Delta\Sigma}$) between
 the plasma source and the event horizon, where $B_{0}=$ constant 
 ($\neq 0$) and the value should be determined at the plasma injection
 point.     
 Some special radii for the MHD flow are given by the field aligned
 parameters. The locations of the light surfaces are given by the  
 $\Omega_F$ value, and the locations of the Alfv\'{e}n radii are 
 given  by the $\Omega_F$ and $\tilde{L}$ values.  In addition, 
 setting $\Gamma=4/3$, the values of $\hat{E}\equiv E/\mu_{c}$ and  
 $\hat{\eta}\equiv 4\pi\mu_{c}\eta /B_0$ with the values of $\Omega_F$
 and $\tilde{L}$ determine the locations of the fast and slow
 magnetosonic points. Then, solution curves $M^2=M^2(r;\dot{\cal{M}})$
 are obtained on the $r$-$M^2$ diagram, and the final parameter
 $\dot{\cal{M}}$ is used to specify an acceptable trans-magnetosonic
 solution for accretion.  
 ( The solution curves also include an outgoing physically acceptable
 solution that can take a different value of $\dot{\cal{M}}$ with the
 accretion solution.  The most remaining solution curves are unphysical. )    
%%\footnote{
%%~~~ NOTE : ~~ for numerical calculation
%% \begin{equation}
%%  (B_0\dot{\cal{M}}) = 
%%   \frac{M^2}{\hat{\eta}} \left(\frac{\mu_{c}}{\mu}\right) 
%%  \left[ \frac{1}{1+N}\left( \frac{\mu}{\mu_{c}}-1 \right) \right]^{N}  
%% \end{equation}
%%}

 By assuming a cold preshock flow 
 ($\dot{\cal{M}}_1 = \dot{\cal{M}}_{\rm pre}=0$), 
 the shock condition (\ref{eq:Well}) can be reduced to
\begin{equation} 
  \left(\frac{ \hat{B_p} }{ \hat{\eta} }\right)^2 M^2_1 
     +\frac{1}{2} \hat{E}^2 f_1^2   
  = \left(\frac{ \hat{B_p} }{ \hat{\eta} }\right)^2 M^2_2 
     +(\hat{\eta} B_0 \dot{\cal M}_2)^{\Gamma-1}  
      \left[\frac{1}{M^2_2}\left(\frac{\mu_2}{\mu_{c}}\right)\right]^\Gamma
     +\frac{1}{2} \hat{E}^2 f_2^2    \ .                \label{eq:sk-tad}
\end{equation}
 This equation refers to the jump of the Mach number across the shock 
 front. When we obtain the two trans-fast (or trans-slow) magnetosonic
 solutions $M^2=M^2_{\rm pre}(r)$ and $M^2=M^2_{\rm post}(r)$ by tuning
 both $\hat{E}$ and $\hat{\eta}$ values with the requirement of 
 $\dot{\cal M}_1 = 0 $, we can apply the solutions to equation
 (\ref{eq:sk-tad}); that is, we can set 
 $M^2_1 = M^2_{\rm pre}(r_{\rm sh})$ and  
 $M^2_2 = M^2_{\rm post}(r_{\rm sh})$.  Then, we obtain the location
 $r=r_{\rm sh}$ satisfying the condition 
 $\dot{\cal{M}}_2 - \dot{\cal{M}}_{\rm post} = 0$.  
 Note that for an acceptable shocked accretion solution we must require  
 the condition $\dot{\cal M}_2 > \dot{\cal M}_1 $, while the other field
 aligned parameters $\Omega_F$, $E$, $L$ and $\eta$ are conserved across
 the shock.

%\placefigure{fig:fast-sk} %---------  Figure~3 --------------------

\begin{figure}[t]%[h]% ------------------------------------- Figure~3
   \epsscale{0.45}
   \plotone{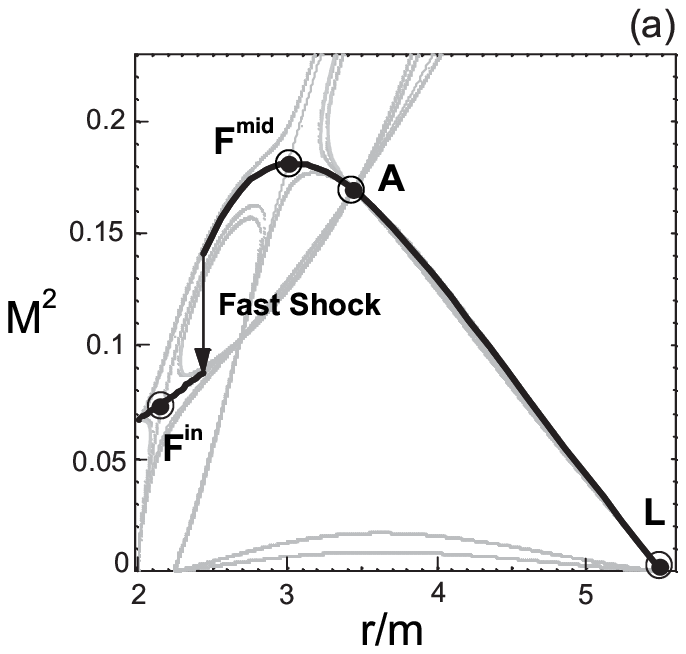}
   \plotone{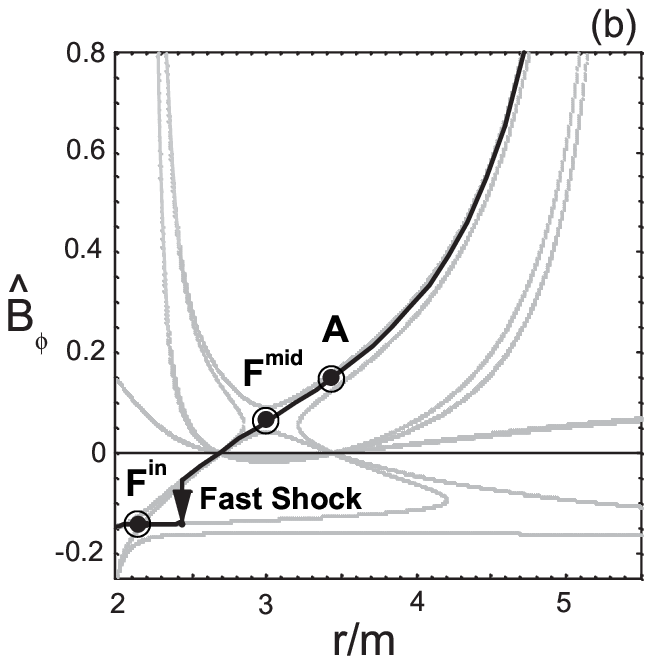}
  \caption{
   The trans-fast magnetosonic accretion solution with a fast
   magnetosonic shock (thick black curves); (a) The square of the Alfv\'{e}n 
   Mach number $M^2$ vs. radius $r/m$ and (b) the toroidal component of
   the magnetic field $\hat{B}_\phi$ ($\equiv B_\phi/B_0$) vs. radius
   $r/m$ relation are shown with $B_0\dot{\cal{M}} = $ constant. After
   passing through the outer-Alfv\'{e}n point (A) and the middle-fast 
   magnetosonic point (F$^{\rm mid}$), the ingoing flow makes the fast
   magnetosonic shock near the event horizon. The postshock flow passes
   through the inner-fast magnetosonic point (F$^{\rm in}$), and then
   falls into the black hole ($r=2m$).  
   The gray curves also show $B_0\dot{\cal{M}} = $ constant, but these
   are unphysical solutions. 
   The flow parameters are given by $\hat{\eta}=0.0108$, $\hat{E}=3.5$, 
   $\tilde{L}/m=4.1$, $m\Omega_F=0.14503$, $\Gamma=4/3$, $a=0$ and
   $\theta=\pi/2$. 
   We obtain $B_0\dot{\cal{M}}_2 = 0.05134$ for the postshock flow, 
   while we give $B_0\dot{\cal{M}}_1 = 0$ for the preshock flow. 
  }
  \label{fig:fast-sk}
\end{figure} %------

 Figure~\ref{fig:fast-sk}a shows the square of the Alfv\'en Mach number
 of the shocked accretion solution vs the radial distance. The solution,  
 which passes through the fast magnetosonic points twice, is represented
 as two $\dot{\cal{M}} = $ constant curves and a vertical arrow (the
 fast magnetosonic shock).   
 The upstream SAF-curve follows the hydro-like solution, although the
 slow magnetosonic point is absent on this diagram because of the cold 
 approximation. Then, at the shock location, it transfers to the
 downstream curve which is the magneto-like solution, and it becomes 
 sub-fast magnetosonic. The shocked flow passes through the fast
 magnetosonic point again to fall into the black hole.  
 Figure~\ref{fig:fast-sk}b shows the toroidal component of the magnetic
 field ($\hat{B}_\phi \equiv B_\phi/B_0$) for the shocked solution as a
 function of radial distance. As the plasma falls inward, the
 trailed-shape of the magnetic field line ($B_\phi<0$) changes to the
 leading-shaped one ($B_\phi>0$) at the Alfv\'{e}n radius (the anchor
 point; not the Alfv\'{e}n point), where $B_\phi=0$.  The strength of
 the postshock toroidal magnetic field $|B_\phi|$ increases across the
 fast magnetosonic shock.     
 In Figure~\ref{fig:fast-sk} the preshock flow solution is plotted as a 
 bold curve starting from the outer light surface with zero-velocity,
 where the plasma rotates toward the $\phi$-direction with the speed of
 light.  So, a realistic flow should start from a location somewhat
 inside the outer light surface.  In the case of (FS-i), the preshock
 solution should connect to the plasma source at some location between
 the outer-Alfv\'{e}n point and the outer light surface. 
 In Figure~\ref{fig:fast-sk}, gray curves also show $\dot{\cal{M}}=$ 
 constant curves, but they do not pass though the magnetosonic points,
 or do not connect to the plasma source, except for the upstream
 $\dot{\cal{M}}=0$ curve of $ r_{H} < r < r_{\rm sh} $. 
 Note that the upstream $\dot{\cal{M}}=0$ curve directly connects the
 plasma source to the event horizon, where the Mach number has a finite
 value at the event horizon.  This curve is also a physically acceptable 
 SAF-solution {\it without} the MHD shock. Here, we cannot discuss which
 of the two accretion solutions; the shocked inflow or the shock-free 
 inflow, is selected as a realistic accretion flow. To answer this
 question, the stability analysis for shocked MHD flows would be
 necessary, but that is beyond the scope of our current paper. In the 
 following, we will go ahead to further explore the shocked MHD 
 accretion flows.

%\placefigure{fig:eta-e} %---------  Figure~4 --------------------

\begin{figure}[t]%[h]% ------------------------------------- Figure~4
   \epsscale{0.45}
   \plotone{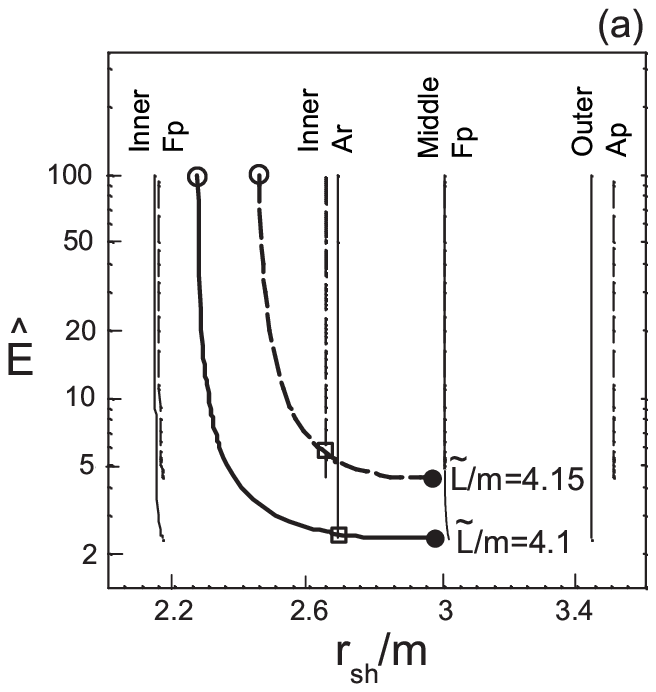}
   \plotone{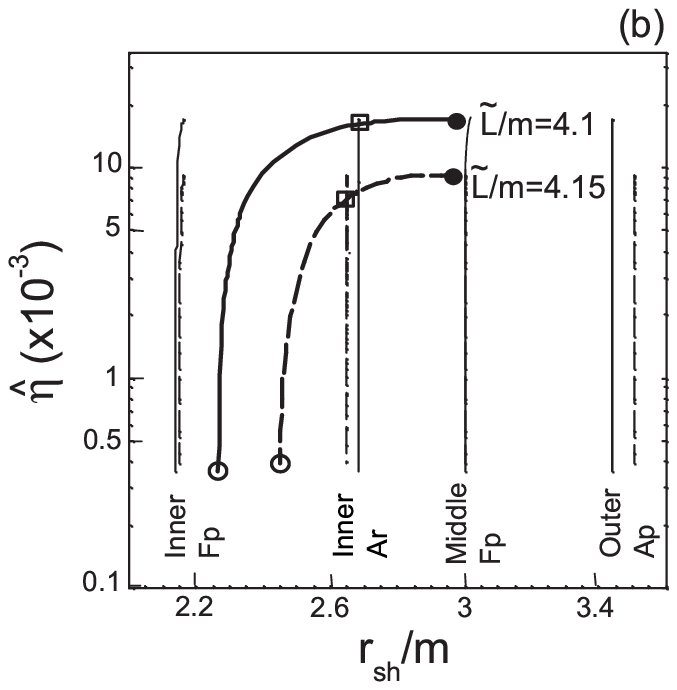}
  \caption{
   The relations between the shock location $r_{\rm sh}/m$ and (a) the
   energy $\hat{E}$, and (b) particle number flux $\hat{\eta}$ for MHD
   shocked accretion flows onto a black hole (FS-i solutions).  
   The flow parameters are given by $\tilde{L}/m=4.1$ (thick solid 
   curves) and $4.15$ (thick broken curves), $m\Omega_F=0.14503$, $a=0$,
   and $\theta=\pi/2$.      
   The branch inside the inner Alfv\'{e}n radius marked by ``$\Box$''
   shows the fast magnetosonic shock, while the branch between the
   radius marked by ``$\bullet$'', which is located inside the
   middle-fast magnetosonic point, and the inner-Alfv\'{e}n radius shows
   the intermediate shock. The curves labeled by ``Fp'' represent the
   radii of the inner/middle fast magnetosonic points. The lines labeled
   by ``Ap'' represent the radius of the Alfv\'{e}n point, and the lines
   with ``Ar'' indicate the Alfv\'{e}n radius. The solid and broken 
   lines/curves correspond to $\tilde{L}/m=4.1$ and $4.15$,
   respectively.   
  }
  \label{fig:eta-e}
\end{figure} %------

\begin{figure}[p]%[h]% ------------------------------------- Figure~5
   \epsscale{0.45}
   \plotone{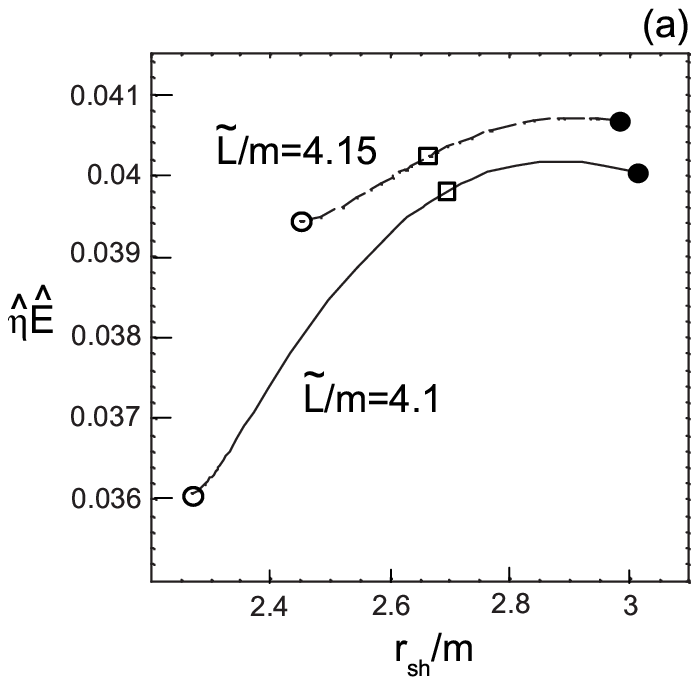}
   \plotone{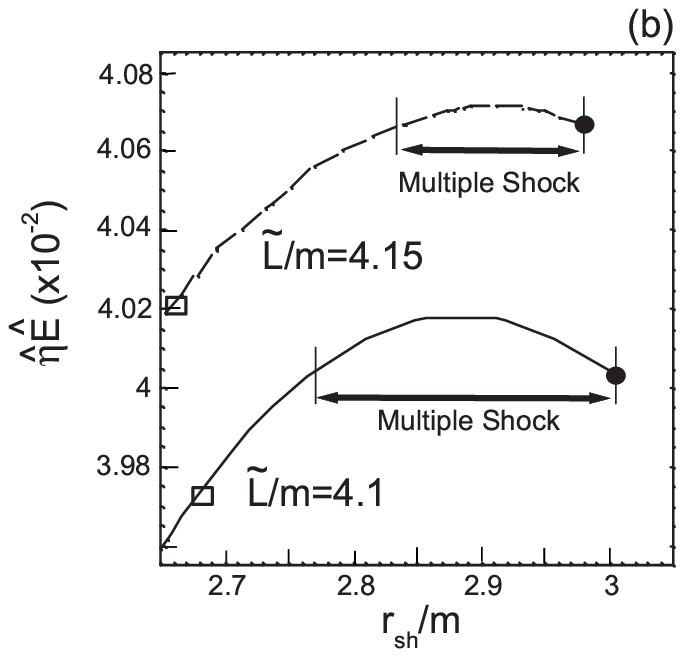}
  \caption{
   The relations between the shock location $r_{\rm sh}/m$ and the
   energy flux $\hat{\eta}\hat{E}$ for a (FS-ii) solution. The multiple
   shock locations are shown in the magnified Right panel (b).  
   The chosen parameter sets are the same as in Fig.~\ref{fig:eta-e}. 
  }
  \label{fig:etaE}
\end{figure}

 The parameter search for the shocked flow solutions is a standard
 approach to  correctly understand the shock formation in a curved
 spacetime (TRFT02) and to estimate the activities of a black hole
 engine. Figure~\ref{fig:eta-e}a shows the relation between the shock
 location $r_{\rm sh}$ and the total energy $E$ for the class (FS-i)
 solutions, which are obtained by solving equation~(\ref{eq:sk-tad})
 with $\dot{\cal{M}}_2-\dot{\cal{M}}_{\rm post}=0$.   
 Figure~\ref{fig:eta-e}b shows the relation between the shock location
 $r_{\rm sh}$ and the particle number flux per the magnetic flux $\eta$.
 The inverse of this quantity exhibits almost the same behavior as $E$.  
 For a physically acceptable shocked MHD accretion solution onto a black 
 hole, the total energy $E$ has the minimum value $E_{\rm min}$ under a
 certain parameter set of $\Omega_F$, $\tilde L$ and $a$. It seems that
 the energy $E$ diverge at $r=r_{\rm sh}^{\rm min}$ (where $\eta E$ has
 a finite value as seen in Fig.~\ref{fig:etaE}), but we stop the
 calculation at $\hat{E}=100$ (marked by $\circ$).  Note that
 $\hat{E}\to\infty$ (and $\eta\to 0$) with $\tilde{L}\Omega_F\to 1$
 corresponds to a force-free limit, where the Alfv\'{e}n radius shifts
 to the light surface; but in this demonstration 
 $\tilde{L}\Omega_F=0.6$; that is, it is not force-free.  
 The possible shock location for physically acceptable shocked accretion
 flows has the minimum radius $r_{\rm sh}^{\rm min}$ and the maximum
 radius $r_{\rm sh}^{\rm max}$ (marked by $\bullet$), where these radius
 are located between the middle- and inner-fast magnetosonic points;
 that is, $r_{\rm F2} < r_{\rm sh}^{\rm min}$ and 
 $ r_{\rm sh}^{\rm max} < r_{\rm F1}$.   
 In Figure~\ref{fig:eta-e}, the radii of the inner-fast and middle-fast
 magnetosonic points weakly depend on the value of $E$ or $\eta$. The
 inner Alfv\'{e}n radius and the outer Alfv\'{e}n point are plotted by
 the vertical lines.

%\placefigure{fig:etaE}    %---------  Figure~5 --------------------

 %------

%\placefigure{fig:mdot-sh} %---------  Figure~6 --------------------

\begin{figure}[p]%[h]% ------------------------------------- Figure~6
   \epsscale{0.45}
   \plotone{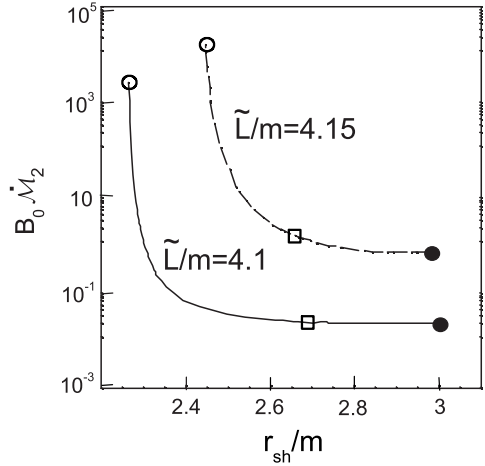}
  \caption{
   The relations between the shock location $r_{\rm sh}/m$ and the 
   entropy generation $B_0\dot{\cal M}_2$ by the shock. 
   The chosen parameter sets are the same as in Fig.~\ref{fig:eta-e}. 
  }
  \label{fig:mdot-sh}
\end{figure} %------

 Figure~\ref{fig:etaE} shows the value of the total energy flux per
 magnetic flux tube $\eta E$ at a fast magnetosonic shock location.  
 At a shock location, the value of $\eta E$ for a shocked flow increases 
 with increasing $\tilde{L}$, while the value of $E$ increases and the
 value of $\eta$ decreases (see Fig.~\ref{fig:eta-e}). Then, the
 fast magnetosonic shock with larger energy flux is obtained for the
 inflow with smaller $E$ (or larger $\eta$) and larger $\tilde{L}$.  
 Figure~\ref{fig:mdot-sh} shows that the relation between the shock
 location and the entropy related mass flow rate per magnetic flux tube
 $\dot{\cal M}_2$ for the postshock flows. For the shock located near
 the inner-fast magnetosonic point the lager entropy generation is
 obtained, while the energy flux becomes small at there.   
 With these flow parameter sets, we may expect a stronger shock
 formation and/or larger energy release from the hot plasma generated 
 by the shock formation. We will discuss this problem later.  
 Note that a maximum value of $\tilde{L}$ exists for the hydro-like
 solution; The hydro-like solution is necessary for the middle-fast
 magnetosonic point to achieve the class (FS-i) solution
 \citep{Takahashi02a}.

 When a certain value of the total energy $E$ (or $\eta E$) is given,
 one or two shock locations are obtained (see Fig.~\ref{fig:etaE}b to
 confirm the {\it multiple-shock}\/ locations). When we choose another
 value of $E$ (or $\eta E$), different values of $\eta$ and 
 $\dot{\cal M}_2$ are determined from the critical conditions at the
 magnetosonic points.  
%% (Note that, when we give the value of $\eta$, the values of remaining
%%  flow parameters $E$ and $\dot{\cal M}$ are consequently determined.) 
 The number of physically acceptable shock location depends on the total
 energy $E$ (or $\eta E$). For example, the number is one for 
 $E_{\ast} \leq E < E_{\rm max}$, and two for 
 $E_{\rm min} \leq E <E_{\ast}$ (a multiple-shock), where $E_{\ast}$ 
 is given at the outermost shock location $r_{\rm sh}^{\rm max}$.  
 When we find the multiple-shock locations for a given parameter set, 
 we should discuss the stability of MHD shocks at each shock location
 to answer the question of which shock location should be selected as 
 an accretion solution.
 However, that task is beyond the scope of the present paper.  An
 important point is that the maximum and minimum values (labeled by
 `max' and `min') exist; that is, the condition for the shock formation
 is limited by some ranges of the parameter values.

\begin{figure}[p]%[h]% ------------------------------------- Figure~7
   \epsscale{0.45}
   \plotone{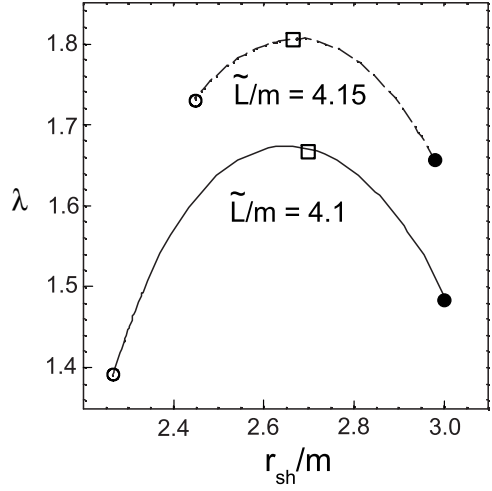}
   \caption{
   The compression ratio $\lambda$ vs.  shock radius relation is
   shown.  
   The chosen parameter sets are the same as in Fig.~\ref{fig:eta-e}. 
   }
  \label{fig:lambda}
\end{figure}

\begin{figure}[p]%[th]% ------------------------------------- Figure~8
   \epsscale{0.45}
   \plotone{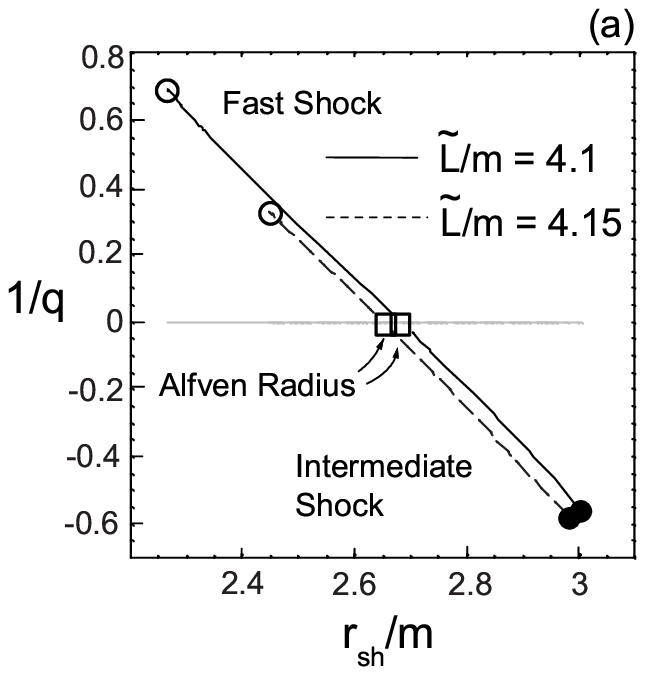}
   \plotone{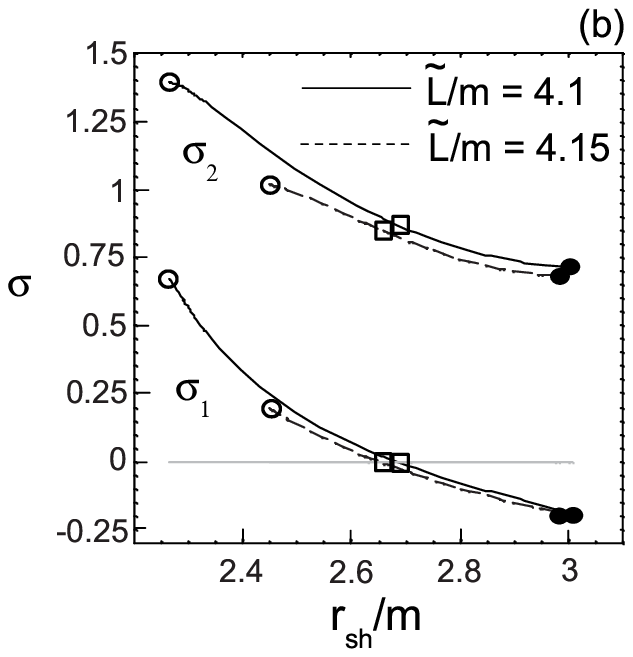}
  \caption{
   (a) The inverse of the toroidal magnetic field ratio across the shock
   $1/q$ vs.  shock radius relation is shown, while   
   (b) the magnetization parameter $\sigma$ for the pre- and post-shock
   inflow vs.  shock radius relation is shown.   
   The chosen parameter sets are the same as in Fig.~\ref{fig:eta-e}. 
  }
  \label{fig:q-sigma}
\end{figure}

\begin{figure}[t]%[h]% ------------------------------------- Figure~9
   \epsscale{0.45}
   \plotone{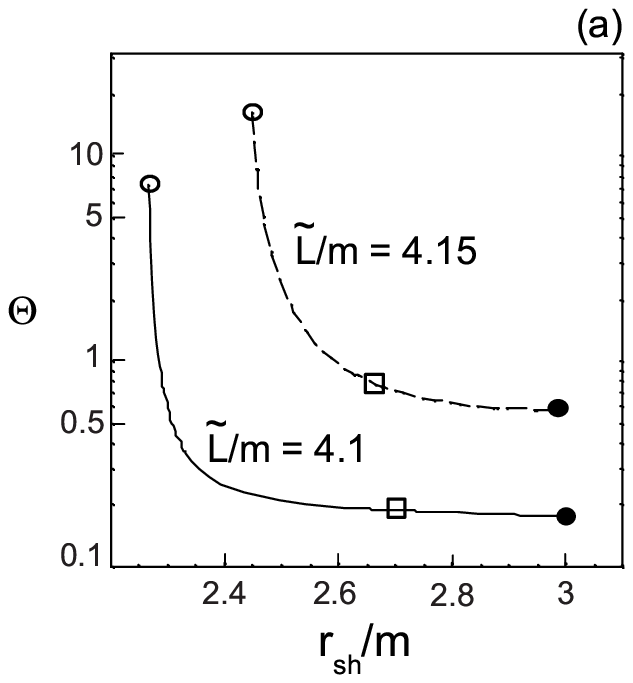} 
   \plotone{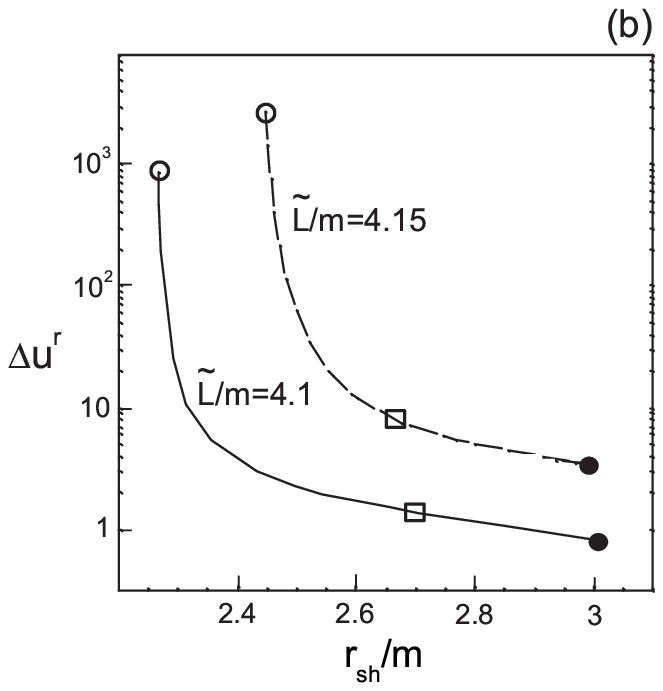} 
  \caption{
  (a) Shows the temperature parameter $\Theta$ vs.  shock radius
   realtion, while    
  (b) shows the jump of the radial four-velocity $\Delta u^r$ vs. 
   shock radius relation.   
  The chosen parameter sets are the same as in Fig.~\ref{fig:eta-e}. 
  }
  \label{fig:temp}
\end{figure}

 If a shock generates between the inner Alfv\'{e}n radius and the
 possible outermost shock radius, which is a segment divided by
 ``$\Box$'' and ``$\bullet$'' in Figs.~\ref{fig:eta-e} -~\ref{fig:temp},   
 the postshock flow becomes sub-Alfv\'{e}nic.  Such a shock is called
 the ``intermediate'' magnetosonic shock (the stability of intermediate
 shocks is discussed by \cite{Hada94}).   
 In the case of an intermediate shock, the post shocked flow must
 pass through both inner Alfv\'en point and inner-fast magnetosonic
 point. In Figures~\ref{fig:eta-e} and~\ref{fig:etaE}, we see that the
 multiple-shock locations are caused in the intermediate shock region.      
 [ For a hot MHD inflow streaming along an area-expanding (diverging)
 magnetic field line of $B_p\propto (1/\sqrt{\Delta\Sigma})r^{-2}$,
 which is converging along ingoing flows, the possible multiple-shock
 locations extends to the fast magnetosonic shock region
 \citep{Goto03}. ]   
 For smaller values of $E$ (or larger $\eta E$), multiple-intermediate
 shock solutions are possible, while an accretion flow with a larger
 value of $E$ (or smaller $\eta E$) produces a fast magnetosonic shock.

%\placefigure{fig:lambda} %---------  Figure~7 --------------------

 %------

 Figure~\ref{fig:lambda} shows the plasma frame compression ratio to be
 $\lambda < 1.7 $ for $\tilde{L}/m=4.1$ and $\lambda < 1.8 $ for 
 $\tilde{L}/m=4.15$; the compression ratio $\xi$ seen by a distant
 observer is somewhat small because the factor $ u^t_{2}/u^t_{1} <1 $
 weakens the shock strength. Note that the Lorentz factor $u^t$ includes
 the effects of strong gravity (gravitational red-shift) and the Lorentz
 boost by the rapid motion of the plasma around the black hole. 
 The strongest shock (the largest $\lambda$ value) is generated around
 the inner Alfv\'{e}n radius,  where the difference in the Alfv\'{e}n
 Mach number between the preshock and postshock trans-fast magnetosonic
 solutions is also the largest. 
 Although the number density $n_1$ of the cold preshock flow is 
 inversely proportional to $M^2_1$, the function $n_1=n_1(r)$ has the 
 minimum value around the middle-fast magnetosonic point.      
%%\footnote{ ~~ NOTE :~~ 
%% \begin{equation}
%%    \tilde{P}_2 - \tilde{P}_1 = \frac{4\pi\eta^2}{U^2}(M^2_1-M^2_2) \ . 
%% \end{equation}
%%}

%\placefigure{fig:q-sigma} %---------  Figure~8 --------------------

 %------

 When the shock is generated just on the inner Alfv\'{e}n radius that 
 is the anchor point for the upstream flow, we see the {\it switch-on
 shock}\/  with  $| 1/q |=0$ (see Fig.\ref{fig:q-sigma}a), where
 $(B_\phi)_1=0$ for the preshock flow and $(B_\phi)_2\neq 0$ for the
 postshock flow.  In the fast magnetosonic shock of the class (FS-i),
 the inner Alfv\'{e}n radius is not the Alfv\'{e}n point for the
 upstream flow solution but it is the Alfv\'{e}n point for the
 downstream flow solution.  The upstream flow becomes hydrodynamical
 ($E=\mu u_t$, $L=-\mu u_\phi$) just on the anchor point. 
 For the positive $q$, as the ratio of the preshock to postshock
 toroidal magnetic fields $ 1/q $ increases in magnitude, the
 magnetization parameter of the postshock flow also increases (see
 Fig.\ref{fig:q-sigma}b). The property $q>1$ %($ 1 > 1/q > 0 $) 
 is a characteristic of the fast magnetosonic shock.    
 The negative $q$ corresponds to the intermediate MHD shock, where the  
 sign of the toroidal component of the magnetic field is reversed. 
 In the class (FS-i) solution, the Poynting flux directs inward
 across the fast magnetosonic shock; that is, the electromagnetic energy
 is carried to the black hole by the preceding rotation of the anchor
 point.  However, for the intermediate MHD shock, the Poynting fluxes
 direct outward (inward) for the preshock (postshock) flows; in the
 intermediate MHD shock the magnetic field line is flipped over at the
 shock normal.

%\placefigure{fig:temp} %---------  Figure~9 --------------------

 %------

 As seen in Fig.~\ref{fig:fast-sk}, when the shock is approaching the
 inner-fast magnetosonic point, the gap between $M^2_{1}$ and $M^2_{2}$
 becomes small. One may expect that only some fraction of kinetic energy
 of the preshock flow converts to thermal energy of the postshock
 flow. In fact, the strength of the shock (the compression ratio
 $\lambda$) is weakened; this means that the ratio of the preshock and
 postshock 4-velocities $(u^r_1/u^r_2)_{\rm sh}$ becomes small. 
 To confirm this, in Figure~\ref{fig:temp}a, the temperature of the
 postshock flow is shown as a function of the shock radius. 
 Figure~\ref{fig:temp}a (see also Fig.~\ref{fig:mdot-sh}), however,
 indicates that the plasma becomes hotter for the fast magnetosonic
 shock generated near the inner-fast magnetosonic point. This is because
 for the shock located near the inner-fast magnetosonic point the
 plasma's energy per rest-mass energy $\hat{E}$ is very large and the
 number flux per magnetic tube $\hat{\eta}$ is small. Furthermore, as
 seen in Figure~\ref{fig:temp}b that shows the jump of the radial
 4-velocity  $\Delta u^r \equiv (u_1^r)_{\rm sh} - (u_1^r)_{\rm sh}$ 
 at the shock location, the gap $\Delta u^r$ becomes large near the 
 inner-fast magnetosonic point even if the ratio 
 $(u^r_1/u^r_2)_{\rm sh}$ is small (where the value of $\lambda$ has a
 small value), and a considerable 
 amount of plasma kinetic energy converts to the thermal energy. 
 The plasma kinetic energy, of course, is converted to the magnetic
 energy by increasing $B_\phi$. However, for the fast magnetosonic shock
 generated near the event horizon, the strength of $(B_\phi)_2$ is   
 $\sim B_{\phi}^{H}~ [ \propto (\omega_{H}-\Omega_F)B^r_{H} ]$, which is  
 determined by the boundary condition at the event horizon (see
 Fig.~\ref{fig:fast-sk}b) and is independent of $u^r_{H}$ and $\hat{E}$
 (and $\hat{\eta}$). We can say that the toroidal magnetic field
 $B_\phi$ of the stationary shocked accretion flow is restricted by the
 general relativistic effect (i.e., the hole's boundary condition),
 although the information of the horizon cannot propagate to the shock
 front. Then, the magnetic energy generation cannot dominate in the
 shock process.  
 Thus, even if the shock is weak, the high temperature postshock inflow
 is obtained. We can also see that the temperature of the postshock
 inflow is rising as the shock location approaches the event
 horizon. So, a cone-like hot plasma column would be built on the event
 horizon. The high-energy emissions are expected from this hot plasma
 column with $ r_{H} < r < r_{\rm sh} $.

%\placefigure{fig:slow} %---------  Figure~10 --------------------

\begin{figure}[t]%[h]% ------------------------------------- Figure~10
   \epsscale{0.45}
   \plotone{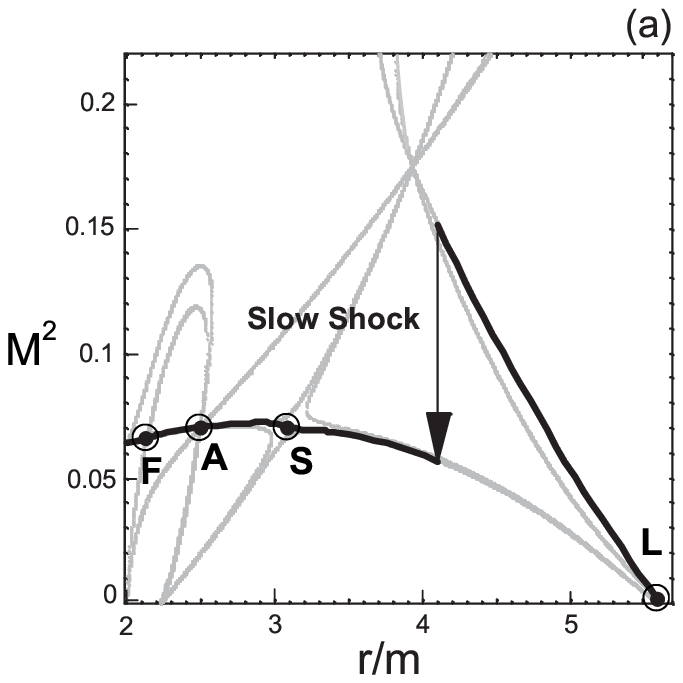}
   \plotone{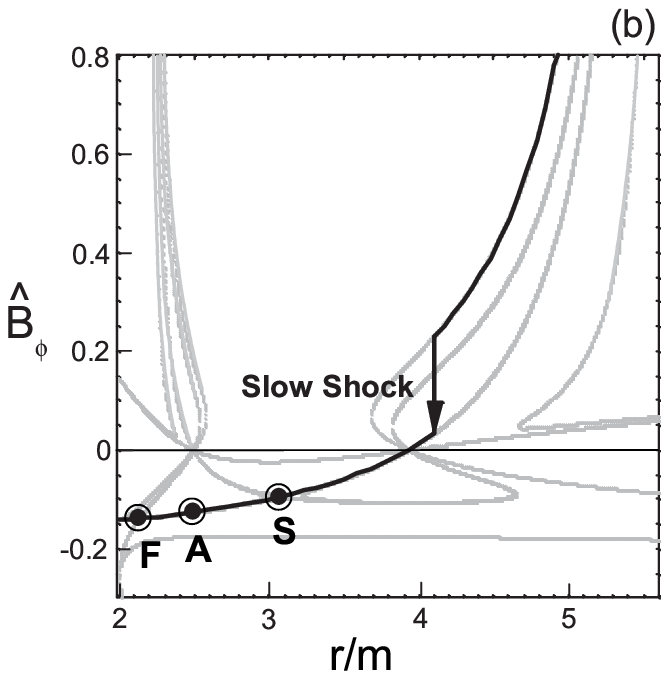}
  \caption{
   A MHD accretion flow with a slow magnetosonic shock is shown:   
   (a) The square of the Alfv\'{e}n Mach number and (b) the toroidal
   magnetic field vs. radius $r/m$ (thick black curves). The preshock
   curve does not connected to the horizon, but it is physically
   acceptable through the shock. 
   The flow parameters are given by $\hat{\eta}=0.00521$, $\hat{E}=7.012$, 
   $\tilde{L}/m=4.5$, $m\Omega_F=0.14303$, $\Gamma=4/3$, $a=0$ and
   $\theta=\pi/2$.    
   We obtain $B_0\dot{\cal{M}}_2 = 3.591$ for the postshock flow, 
   while we give $B_0\dot{\cal{M}}_1 = 0$ for the preshock flow. 
  }
  \label{fig:slow}
\end{figure} %------

 Figure~\ref{fig:slow} shows an accreting flow with the slow
 magnetosonic shock. Here, the upstream flow solution is cold, so that
 the injected plasma from a plasma source is super-slow magnetosonic.
 (Note that the upstream super-slow magnetosonic solution does not
 necessarily have to pass through the middle-fast magnetosonic point.)   
 After the slow magnetosonic shock, the ingoing flow becomes hot, and
 passes through the slow magnetosonic point, the Alfv\'{e}n point and
 the fast magnetosonic point in this order, and then falls into a black
 hole.  Although, here the slow magnetosonic solution is generated near
 the outer Alfv\'{e}n radius, the jump of the Mach number of the slow
 magnetosonic shock is larger than that of the fast magnetosonic shock
 (see Fig.~\ref{fig:fast-sk}); that is, the compression ratio becomes
 higher.   
 The preshock toroidal component of the magnetic field decreases with
 decreasing radius, and at the slow magnetosonic shock it
 decreases catastrophically. After the slow magnetosonic shock, the
 trailed-shape of the magnetic field line changes to the leading one. 
 The detailed properties of the slow magnetosonic shocks are discussed
 by TRFT02 and  %[cut] Rilett 
 Takahashi et al. (in preparation).

\section{ Summary and Conclusion }

 We have formulated the MHD shock conditions in Kerr geometry and  %[add]
 have discussed MHD accretion flows onto a 
 non-rotating  %[add]
 black hole with
 magnetosonic shock formation. We find that a very hot plasma region 
 can form close to the black hole. To realize these flows,  solutions
 with multiple magnetosonic points are necessary, where the critical
 conditions at the magnetosonic points are required for the acceptable
 set of the field-aligned parameters.  Furthermore, the MHD shock must
 be located somewhere between the two magnetosonic points, where the
 shocked plasma must satisfy the jump conditions.  
 Although the flow parameters should be specified at a plasma injection 
 source (e.g., the  surface of an accretion disk or its corona, etc), 
 these parameters for the acceptable MHD accretion flows are specified
 at some points located near the event horizon by the above conditions.
 Then, the shocked accretion phenomena would give us information about
 the plasma sources and the magnetosphere (the inflowing region) in a
 curved spacetime.

 A strong MHD shock with large fluid compression is obtained when the
 Mach number gap between the preshock and postshock flows is large; such 
 a shock generates around the inner-Alfv\'{e}n radius.  We find that
 this situation is realized through the transition from the hydro-like
 preshock flow to the magneto-like postshock flow (type FS-i shocked MHD 
 accretion flow).     
 In the case of a rapidly rotating black hole with $0<\Omega_F<\omega_H$, 
 this situation with the hydro-like and magneto-like solutions is lost. 
 However, this is possible for the type (FS-ii) shocked MHD accretion
 flows. In that case, one may not be able to expect strong MHD
 shock formation with a large compression ratio because the Mach number 
 jump is not so large as in the (FS-i) case. However, a very hot plasma
 that would be caused by the entropy generation at the shock with  
 $r_{\rm sh} \sim r_{\rm sh}^{\rm min}$ would be expected as in
 the case of (FS-i) shocked accretion flows.

 The Lorentz factor of relativistic jets observed in AGNs is $\sim$ 
 10 -- 100. Therefore, we expect, as an origin of the jet, that the
 ejected plasma wind from the disk surface has the total energy of 
 $\hat{E}\sim$ 10 -- 100.  The ejected high-energy plasma streams along 
 the magnetic field lines in the magnetosphere, and forms both outgoing
 winds and ingoing winds, where the separatrix surface, which separates 
 inflow and outflow regions, depends on the magnetic field distribution
 in the magnetosphere.  Although the outgoing wind outside the separatrix 
 surface is accelerated and would make a relativistic jet, some part of 
 the ejected wind inside the separatrix surface streams toward the black
 hole because of its strong gravity. Such a wind can produce a MHD shock
 discussed in this paper. The shocked plasma with $\hat{E}\sim$ 10 --
 100 will then make a very hot plasma region.    
 Thus, we can expect that a very hot plasma region is located near the
 event horizon as a source of high-energy radiation (X- and $\gamma$-ray 
 emissions). Of course, most radiation fluxes would fall into the black
 hole by the gravitational lens effect, and the energy of the outgoing
 radiation is lost by the gravitational red-shift effect. However, a
 huge energy radiation generated by the shock would modify the 
 distributions of the plasma and magnetic field geometry around the
 black hole.   
 For example, some kind of plasma instabilities or dynamical phenomena, 
 which will include the information on the strong gravity, may be caused
 by the strong radiation. Such plasma phenomena taking place between the
 shock and the inner-fast magnetosonic point can propagate outward, and
 may influence the upstrem flow and further the plasma source.  
 In a magnetized disk--black hole system, ``hyperaccretion'' onto a slowly
 rotating black hole is proposed as a model for short gamma-ray bursts
 \citep{PO01}. If a very hot plasma region by the MHD shock is generated
 along the hyperaccretion flow, a fraction of accreting plasma may be
 blown away as a GRB jet by some plasma instabilities or dynamical 
 phenomena.

 When we are interested in a dense accretion flow (with a larger $\eta$
 value) that leaves the disk surface without large initial velocity,  
 the expected energy would be $\hat{E} \leq 1$. Although we have
 explored MHD shocks for larger $\hat{E}$, we also tried to search for
 the minimum energy of shocked MHD accretion flows (FS-i) under wide 
 ranges of flow parameters (including the variations of $\tilde{L}$,
 $\Omega_F$ and $\theta$ values).  However, we did not find a MHD shock
 solution for $\hat{E} \leq 1$ in our limited parameter search.  
 More systematic analysis will be presented in our subsequent paper
 (Fukumura et al. in preparation).

 In the case of quiet black hole accretion system ($\hat{E}\leq 1$), no 
 MHD shock may be obtained near the equator. However, the MHD shock can
 be generated along a disk--black hole magnetic field line, and the hot
 plasma region can form in the high-latitude region of the black
 hole. Then, we expect that as a signature of AGN activities X- and
 $\gamma$-ray emissions from the off-equatorial shocked hot plasma are
 observed directly to us and would be capable of locally illuminating
 the underlying accretion disk. This could photoionize iron atoms in the 
 disk, causing subsequent iron fluorescence observed in many Seyfert
 nuclei and Galactic BH candidates. Our MHD shock model thus can be a
 probable local X-ray radiation source in these systems. 
 Although the estimate of the emergent X-ray spectrum (at the shock) is
 important, it is beyond  the scope of our current investigation.

%--------------------------------

\acknowledgments

We would like to thank Akira Tomimatsu for his helpful comments. 
This work was supported in part by the Grants-in-Aid of the Ministry of 
Education, Culture, Sports, Science and Technology of Japan
(17030006,M.T).

\appendix

\section{Shock Condition in Plasma Comoving Frame}

 In this paper, we use the expression with the lab-frame magnetic field 
 $B^\alpha$ to analysis the MHD shock conditions. Here, we show the
 relations between the lab-frame magnetic field and the plasma comoving
 frame magnetic field.  

 The magnetic and electric fields in the plasma comoving frame are 
 defined by $ b_\alpha \equiv {^\ast F}_{\alpha\beta} u^\alpha $ 
 and $e_\alpha \equiv F_{\alpha\beta}u^\beta$, which satisfy 
 $e_\alpha=0$ for the ideal MHD plasma. Then, the homogeneous Maxwell
 equation   %% $F_{[\alpha\beta;\gamma]}=0$ 
 can become $(u^\alpha b^\beta-u^\beta b^\alpha)_{;\alpha}=0$, which
 means the magnetic flux conservation, and equation (\ref{eq:emt})
 becomes  
\begin{equation}
 T^{\mu\nu} = w u^\alpha u^\beta - \tilde{P} g^{\alpha\beta} 
             -\frac{1}{4\pi}b^\alpha b^\beta   \ ,   \label{eq:emt2}
\end{equation}
 where $w \equiv n\mu+(b^2/4\pi)$, $\tilde{P}\equiv P + (b^2/8\pi)$ 
 and $b^2\equiv -b^\alpha b_\alpha$. 
 The lab-field $B^\alpha$ denoted in the Boyer-Lindquist coordinates 
 are related to the comoving magnetic field $b^\alpha$ by 
\begin{eqnarray}
 B_r &=& \sqrt{-g} F^{\phi\theta} \ = \ G_t b_r / \hat{e}       \ ,  \\
 B_\theta &=& \sqrt{-g} F^{r\phi} \ = \ G_t b_\theta / \hat{e} \ ,  \\
 B_\phi   &=& \sqrt{-g} F^{\theta r} \ 
           = \ (G_t b_\phi - G_\phi b_t ) / \hat{e}       \ , 
\end{eqnarray}
 where $ \hat{e} \equiv (E-\Omega_F L)/\mu $.
 From the shock conditions, the following scalar $U$ and two vectors 
 remain unchanged across the shock \citep{Lichnerowicz67} : 
 \begin{eqnarray}
   U        &\equiv& n u_\perp \ ,   \\ 
   V^\alpha &\equiv& b_\perp u^\alpha - u_\perp b^\alpha \ , \\ 
   W^\alpha &\equiv& u_\perp w u^\alpha 
                  - \tilde{P}\ell^\alpha - (1/4\pi)b_\perp b^\alpha  \ , 
 \end{eqnarray}
 where $u_\perp \equiv u^\alpha \ell_\alpha$ and 
 $b_\perp \equiv b^\alpha \ell_\alpha$ (the index ``$\perp$'' indicates
 the normal component of the vector to the shock front).  Note that 
 $V^\alpha\ell_\alpha=0$. 
 Furthermore, from the product $V^\alpha W_\alpha$ we obtain the
 relation $[\mu b_\perp]^{1}_{2} = 0$, which is equivalent to 
 $[B_\perp]^{1}_{2} = 0$, 
 where $B_\perp\equiv B^\alpha\ell_\alpha$.   
 Thus, the qantity $B_\perp$ ( or $\mu b_\perp$) in a curved-space 
 is a shock invariant. Note that the normal component of the
 comoving-frame magnetic field changes across the shock, while in the
 non-relativistic limit we find $\mu=\mu_{c}=m_{\rm part}$, so that
 $b_\perp$ remains unchanged.

%\clearpage %==========================================================

%=========================================================================

\end{document}